\documentclass[12pt]{article}
\usepackage[a4paper, total={6in, 8in}]{geometry}

\usepackage[utf8]{inputenc}
\usepackage[english]{babel}
\usepackage{tabularx}
\usepackage{graphics}
\usepackage[pdftex]{graphicx}
\usepackage{psfrag}
\usepackage{epsfig}
\usepackage{subfig}
\usepackage{amsmath}
\usepackage{mathtools}
\usepackage{amssymb}
\usepackage{bbold}
\usepackage{bbm}
\usepackage{bm}
\usepackage{setspace}
\usepackage{rotating}
\usepackage{colortbl}
\usepackage{slashed}
\usepackage{braket}
\usepackage{lineno}
\makeatletter
\usepackage{textcomp}
\usepackage[usenames,dvipsnames]{xcolor}
\usepackage{relsize}
\usepackage{cite}
\usepackage{float}
\usepackage{simpler-wick}
\usepackage{enumerate}
\usepackage{verbatim}
\usepackage{fancyhdr}
\usepackage{multirow}
\usepackage{arydshln}
\usepackage{enumitem}
\usepackage{hyperref}
\hypersetup{colorlinks,citecolor=nicegreen,linkcolor=niceblue}
\hypersetup{colorlinks=true}
\usepackage{pifont}
\usepackage{physics}
\usepackage{cancel}
\usepackage{setspace}

\allowdisplaybreaks

\newcommand{\bea}{\begin{eqnarray}}
\newcommand{\eea}{\end{eqnarray}}
\newcommand{\beq}{\begin{equation}}
\newcommand{\eeq}{\end{equation}}
\newcommand{\ec}{\end{center}}
\newcommand{\bc}{\begin{center}}

\newcommand{\pdir}{p\kern -5.2pt\raise 0.2ex\hbox {/}}

\newcommand{\vdir}{v\kern -5.75pt\raise 0.15ex\hbox {/}}
\newcommand{\kdir}{k\kern -5.75pt\raise 0.15ex\hbox {/}}
\newcommand{\epsdir}{\epsilon\kern -5.0pt\raise 0.15ex\hbox {/}}
\newcommand{\bvdir}{\bar{v}\kern -5.75pt\raise 0.15ex\hbox {/}}
\newcommand{\Ddir}{D\kern -7.75pt\raise 0.20ex\hbox {/}}
\newcommand{\Adir}{A\kern -7.75pt\raise 0.20ex\hbox {/}}
\newcommand{\ldir}{l\kern -5.0pt\raise 0.2ex\hbox{/}}
\newcommand{\varepsdir}{\varepsilon\kern -5.5pt\raise 0.15ex\hbox{/}}

\newcommand{\PS}{\mathcal{P}}
\newcommand{\VE}{\mathcal{V}}

\newcommand{\nn}{\nonumber}

\makeatother

\definecolor{niceblue}{rgb}{0.15,0.15,0.6}
\definecolor{nicegreen}{rgb}{0.1,0.5,0.1}
\definecolor{Red}{rgb}{1.,0.,0.}

\begin{document}
\unitlength = 1mm

\thispagestyle{empty} 

\unitlength = 1mm

\thispagestyle{empty} 

\begin{center}
\vskip 3.4cm\par
{\par\centering \textbf{\Large\bf ALP production in weak mesonic decays}}
\vskip 1.2cm\par
{\scalebox{.85}{\par\centering \large  
\sc  A.W.M.~Guerrera$^{a}$ and S.~Rigolin$^{a,b}$} \\
{\par\centering \vskip 0.7 cm\par}
{\par\centering \vskip 0.25 cm\par}
{\sl $^a$ Istituto Nazionale Fisica Nucleare, Sezione di Padova \\ 
     $^b$ Dipartamento di Fisica e Astronomia ``G.~Galilei", \\ Universit\`a degli Studi di Padova, I-35131 Padova, Italy } \\
{\vskip 1.65cm\par}}

\end{center}

\vskip 0.85cm
\begin{abstract}
Axion--Like--Particles are among the most economical and well motivated extensions of the Standard Model. In this work ALP 
production from hadronic and leptonic meson decays are studied. The hadronization part of these decay amplitudes have been 
obtained using Brodsky--Lepage method or LQCD, at needs. In particular, the general expressions for ALP emission in mesonic 
s-- and t--channel tree--level processes are thoroughly discussed, for pseudoscalar and vector mesons. Accordingly, the calculation 
of the decay amplitudes for $M_I\to M_F \,a$ and $M\to \ell \nu a$  are presented. Finally, bounds on the (low--energy effective Lagrangian) 
ALP--fermion couplings are derived, from present and future flavor experiments.
\end{abstract}
\newpage
\setcounter{page}{1}
\setcounter{footnote}{0}
\setcounter{equation}{0}
\noindent

\renewcommand{\thefootnote}{\arabic{footnote}}

\setcounter{footnote}{0}


\section{Introduction}

Light pseudoscalar particles are a common feature of many extensions of the Standard Model (SM) of particle physics. These can 
be naturally introduced in beyond SM (BSM) scenarios, following the QCD axion paradigm \cite{Peccei:1977hh,Peccei1977,Wilczek:1977pj,
Weinberg:1977ma}, as pseudo Nambu-Goldstone bosons (pGBs) of a global $U(1)_{PQ}$ symmetry, non--linearly realized, 
spontaneously broken at a scale $f_a\gg v,$ where $v$ is the Higgs VEV. The main difference between the QCD axion and an 
Axion Like Particle (ALP) lies in abandoning the requirement that the only explicit breaking of the $U(1)_{PQ}$ symmetry arises 
from non--perturbative QCD effects \cite{Weinberg:1977ma}, imposing the well known relation $m_a f_a\approx m_\pi f_\pi$.
Therefore, allowing the ALP mass $m_a$ and the PQ symmetry breaking scale $f_a$ to be independent parameters gives rise to an 
abundance of scenarios populated by scalar singlets under the SM group, not necessarily tied to the solution of the Strong 
CP problem \cite{PhysRevD.19.2227,CREWTHER1979123,Pospelov:2005pr,RevModPhys.82.557,PhysRevLett.97.131801}. 
Notable BSM theories that include light singlet scalar fields are: string theory models \cite{Cicoli:2013ana}, familons 
\cite{PhysRevLett.49.1549,Feng:1997tn}, flaxions \cite{Calibbi:2016hwq,Ema:2016ops} and relaxions \cite{Graham:2015cka}. 
ALPs, regardless the naturalness problem they are called to solve, represent compelling candidates for explaining DM abundance 
in our Universe \cite{Preskill:1982cy,Abbott:1982af,Dine:1982ah}. Whatever is the scenario considered, due to the pGB origin of 
the ALP, it may be plausible that the first hints of New Physics (NP) at the $f_a$ scale could be hidden in low--energy observables. 

The most general CP conserving effective Lagrangian, including operators up to dimension five \cite{Georgi:1986df}, and 
describing ALP interactions with SM fermions and gauge bosons, is given by:
\beq
\delta\mathcal{L}^a=-\frac{\partial_\mu a}{2f_a} \bar{f}\gamma^\mu(C_V+C_A\gamma^5)f - 
                    \frac{\partial_\mu a}{f_a}\sum_Xc_XX_{\mu\nu}^a\tilde{X}^{\mu\nu a},
\label{eq:General_L}
\eeq
where $C_V$ and $C_A$ are hermitian matrices in flavor space, $a$ is the ALP field, $f$ are the SM fermions and $X_{\mu\nu}^a$ 
indicates any SM gauge boson field strength, with $\tilde{X}^{\mu\nu a}\equiv X_{\alpha\beta}^a\epsilon^{\alpha\beta\mu\nu}/2$. 
Note that the flavor conserving diagonal vector couplings are identically zero up to a shift in the electroweak anomalous coupling.
Following most of the literature \cite{Aditya:2012ay,Gavela:2019wzg,Merlo:2019anv,MartinCamalich:2020dfe,Guerrera:2021yss,
Gallo:2021ame}, a low--energy CP and flavor conserving effective Lagrangian for ALP-fermion interactions can be introduced: 
\beq
\delta\mathcal{L}^{a}_{\mathrm{ferm}} = -\frac{\partial_\mu a}{2f_a}c_i\,\bar{f}_i\gamma^\mu\gamma^5 f_i = 
                          i \frac{a}{f_a}  m_{i} \,\bar{f}_i\gamma^5c_if_i.
\label{eq:lag_def_c}
\eeq
The index $i$ extends over all the fermions but the neutrinos, assumed to be massless, with $c_i$ real, but not universal, 
ALP-fermions couplings. With the Lagrangian of Eq.~(\ref{eq:lag_def_c}) all flavor-violating effects will be loop induced 
and CKM suppressed,\footnote{In more general frameworks flavor violating couplings can be introduced at tree level 
\cite{Feng:1997tn,Bauer:2020jbp,MartinCamalich:2020dfe}, and limits on these parameters can be simply recovered by removing 
the loop factors and the CKM suppression.} in the spirit of the Minimal Flavor Violation (MFV) ansatz \cite{DAmbrosio:2002vsn}. 

Most of the attention in the past has been devoted in constraining ALPs couplings with gauge bosons, mainly photons 
and gluons. In particular, very stringent bounds on $c_\gamma$ can be obtained from astrophysical searches: helioscopes 
\cite{PhysRevLett.51.1415,Zioutas:1998cc,Irastorza_2011,Irastorza:1567109}, haloscopes \cite{PhysRevLett.117.141801,
PhysRevLett.122.121802,PhysRevLett.118.091801,Semertzidis:2019gkj}, anomalies in stellar evolution \cite{Ayala:2014pea,
Straniero:316736} or helioseismology \cite{Vinyoles_2015}. ALP-fermions couplings can be studied in astro--particles/DM 
experiments like XENON \cite{XENON100:2014csq,PhysRevLett.118.261301} or CASPEr \cite{PhysRevX.4.021030} and ARIADNE 
\cite{PhysRevLett.113.161801} or using astrophysical data, like for example supernova $\gamma$--ray emission \cite{Payez_2015}. 
All these searches are, however, limited to very light ALP masses, rarely heavier than few hundreds of eV, and, moreover, they 
can only bound first generation ALP-fermion couplings. Therefore, terrestrial beam experiments are complementary in the 
exploration of the ALP couplings parameter space, and, among them, flavor factories are very likely the most promising ones. 

Flavor physics experiments have received more and more attention from the phenomenological community \cite{Izaguirre:2016dfi,
Gavela:2019wzg,MartinCamalich:2020dfe,Bauer:2021mvw,Guerrera:2021yss,Gallo:2021ame}. Strong limits on the ALP couplings in 
Eq.~(\ref{eq:General_L}) can be derived, for example, through the study of the $K\to\pi a$ decay. In large part of the literature 
a flavor universal ALP--fermion coupling, often dubbed $c_{a\Phi}$, is assumed. In this scenario, the $K\to\pi a$ amplitude is 
penguin dominated\footnote{It has to be recalled that this strong bound mainly arises from the top-penguin diagrams, due to 
the large top--mass enhancement.} and one can bounds $c_W$ and $c_{a\Phi}$ at the level of $10^{-3}$ for $f_a=1$ TeV 
\cite{Gavela:2019wzg}. 
However, several models have been introduced where large hierarchies between axion couplings \cite{Choi:2014rja,Kaplan:2015fuy,
Giudice:2016yja,Farina:2016tgd} are naturally produced. It is then of foremost phenomenological relevance to scan the ALP 
couplings parameter space following a less unbiased approach and to identify case by case which limits can be extracted from 
a given experiment on each independent ALP-fermion coupling. For example, in a non--universal ALP--fermion coupling scenario, 
the strongest limit from the $K \to \pi a$ decay is obviously obtained for the ALP-top coupling, $c_t \approx  c_{a\Phi} 
\lesssim 10^{-3}$ for $f_a=1$ TeV, due to the penguin top--mass enhancing. However, being the charm penguin contribution to the 
$K \to \pi a$ roughly $10$\% of the top one, an independent bound on the ALP-charm coupling, $c_c \lesssim 10^{-2}$, can be 
derived, assuming all the other ALP couplings vanishing. Moreover, as noticed by \cite{Guerrera:2021yss}, the $K\to\pi a$ decays 
can also be mediated by tree--level diagrams with a $W^{\pm}$--boson exchanged in the s--channel (t--channel) for charged (neutral) 
$K$ decays. These diagrams contribute at the 1\% level to the total decay amplitude and therefore one can extract independent 
limits on the ALP-lighter quark couplings, $c_{u,d,s} \lesssim 10^{-1}$ for $f_a=1$ TeV, for most of the kinematically allowed 
$m_a$ range. 

However, one of the main obstacles in calculating hadronic observables is to deal with the associated non--perturbative matrix 
element. In treating transitions mediated by local operators, like for example penguin contributions with heavy virtual particles 
in the loop, one can make use of the available Lattice QCD results \cite{Carrasco:2016kpy}. Conversely, to compute products of 
bi--local operators mediated by virtual light states, alternative methods, like for example the Brodsky--Lepage technique 
\cite{Lepage:1980fj,Brodsky:1981rp,Szczepaniak:1990dt}, have to be used \cite{Guerrera:2021yss}. Only when the calculation of 
all these different contributions is done explicitly, one can fairly compare the sensitivity reach on ALP-fermion couplings 
of the different mesonic decay channels.  

From the previous discussion one realizes immediately that, differently from what naively expected, present experiments can 
already provide a full set of constraints on the possible ALP--fermion flavor structures, therefore motivating a thorough study 
of mesonic ALPs rare decays. In this work two wide classes of mesonic decays in ALP are considered: $i)$ the mesonic decays 
$M_I\to M_F a$, with $M_I$ and $M_F$ being either a pseudoscalar or a vector meson and $ii)$ the leptonic meson decay 
processes, $M\to \ell\nu a$. 
In all these processes an ``invisible'' ALP is assumed, i.e. the ALP lifetime is sufficiently long for escaping the detector 
(typically $\tau_a \ge 100$ ps) or alternatively the ALP is mainly decaying in a, not better specified, invisible sector. 

The flagship process for ALP searches in flavor transitions is undoubtedly the $K\to \pi + \cancel{E}$ signature, studied 
at NA62 \cite{NA62:2020pwi,CortinaGil:2020fcx,CortinaGil:2021nts,CortinaGil:2021gga} and KOTO \cite{Ahn:2018mvc}, that arises 
from a $s\to d\, a$ transition at the quark level. Kaon physics is thus in the spotlight for probing ALP couplings in the 
KeV to hundreds of MeV ALP--mass region. $B$--factories have also a fundamental impact in limiting the ALP--fermion coupling 
parameter space. BaBar, Belle \cite{Masso:1995tw,Bevan:2014iga,Dolan:2017osp,Belle:2017oht,Kou:2018nap,CidVidal:2018blh,
deNiverville:2018hrc,Dattola:2021cmw} and LHCb \cite{Aaij:2015tna,Aaij:2016qsm} are carefully analyzing visible and invisible 
signatures of $b$--meson decays. For example, Belle experiment had conducted searches for $B\to M_F \bar\nu\nu$ for many 
different mesonic final states $M_F$, testing ALP masses up to a few GeV. Conversely, measurements of $D$ mesons decays with 
final state composed of a meson and missing energy are missing at the moment \cite{ParticleDataGroup:2020ssz}. Another class 
of interesting processes for probing ALP--fermion physics is represented by decays with a mono--$\gamma +\cancel{E}$ signature. 
Flavor conserving $\Upsilon$ resonant searches exploiting decays such as $\Upsilon(nS) \to \Upsilon(1S)\pi^+\pi^-$ can be 
used to directly probe $\Upsilon(1S) \to \gamma +\cancel{E}$ decays \cite{BaBar:2009gco}.

The paper is organized as follows: in Sec.~\ref{sec:hadronization} a general discussion on the hadronization techniques 
needed for calculating  pseudoscalar and vector meson decays in ALP is presented. 
In Sec.~\ref{Sec:mesonic} the mesonic decay amplitudes in ALPs needed for 
calculating hadronic and leptonic meson decays in ALP are derived in general. Many of these amplitudes have been 
calculated here for the first time. Useful phenomenological approximation are discussed for each channel.
Section~\ref{sec:invisible_bounds_ff} is devoted to describe the phenomenological impact 
of ALP emission in weak--induced meson decays, assuming an invisible ALP in the final state. All the relevant charged and neutral 
meson hadronic decays in ALP are estimated at tree and/or at one--loop (penguin) level. Derived bounds on ALP--fermion couplings 
from hadronic and leptonic meson decays are then discussed and a complete summary of the present situation is shown. 
Finally, for completeness, in Appendix \ref{sec:FlavourViolating} exclusion bounds on flavor changing ALP-fermion parameters 
are presented, for two different scenarios.



\section{Mesonic Hadronization}
\label{sec:hadronization}

Despite the fact that in their rest frame mesons and baryons are complex non-static objects, they interact with highly 
relativistic particles mostly through their valence quark content, justifying a simplified description of these non--perturbative 
states \cite{Lepage:1980fj,Brodsky:1981rp,Szczepaniak:1990dt,EFREMOV1980245,Sterman:1997sx}. In the case at hand, the relevant 
observation is that the final state particles, when recoiling back to back must be emitted in a highly relativistic state. Therefore, 
the strong quantum effects that bind the meson constituents appear highly time-dilated, and the partonic content looks frozen, 
to the light escaping particle. For relative speeds near the speed of light the two recoiling particles are in contact for 
a very short time, decreasing as $(1-v^2/c^2)^{1/2}$. The relevant interactions can then only happen on small time scales and 
distances, relatively to typical mesonic masses and sizes, where QCD is perturbative. As such, the short--distance and the 
long--distance dynamics will have practically no interference. This incoherence between soft and hard physics implies that each 
meson, during the entire interaction with a highly relativistic particle, can be approximated with its partonic structure  
allowing for a significant simplification. 
\begin{figure}[!t]
\center
\includegraphics[scale=0.23]{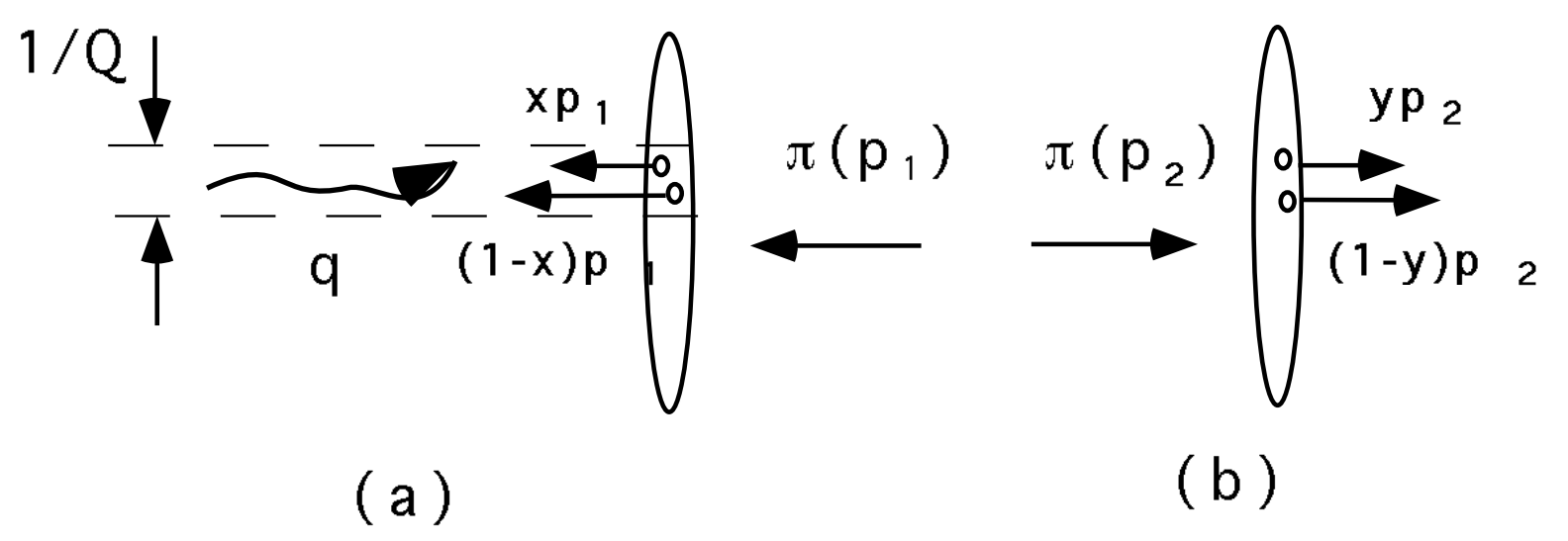}
\caption{The valence quark state of a pion interacting with an external current carrying momentum $\mathcal{Q}$. The valence 
couple must be localized in $1/\mathcal{Q}$ in the transverse direction while in the longitudinal one Lorentz contraction assures 
the partons to be close. The picture is taken from \cite{Sterman:1997sx}.}
\label{fig:pion_picture} 
\end{figure}
The long--time dynamic of the initial state is described by the valence quarks Distribution Amplitude (DA) function, $\phi_I(x)$, 
representing the probability of finding the valence quarks of the incoming meson with a certain meson momentum fraction $x$. On the 
other side, the short--distance interaction between the meson quarks and the ALP is described by the hard--scattering amplitude 
$\Gamma (x,y,\mathcal{Q},\mu)$. 
Eventually, at a later time, quarks reform the outgoing meson, again described via a DA function, $\phi_F(x)$. The total amplitude 
for the process at hand can then be written as a convolution of the three probabilities: 
\beq
\bra{M_F}\Gamma\ket{M_I}=\int dx \, dy\Big (\phi^*_F(y,\mu) \otimes \,\Gamma(x,y,\mathcal{Q},\mu)\,\otimes
                 \phi_I(x,\mu)\Big)\Big(1+\mathcal{O}(m_q^2/\mathcal{Q}^2)\Big)
\label{eq:amplitude_fact}
\eeq
where $m_q$ is the mass of the lightest quark of the meson and $\Gamma(x,y,\mathcal{Q},\mu)$ is to be expanded perturbatively. 
In Eq.~(\ref{eq:amplitude_fact}) $\mathcal{Q}$ is the exchanged momentum, $x,y$ indicate the fraction of momentum carried by 
the heaviest parton of the initial and final meson respectively and $\mu$ is the renormalization scale. A natural choice is 
$\mathcal{Q}=\mu$, making the perturbative calculation consistent as long as $\alpha_{s}(\mathcal{Q})$ is perturbative. The length 
associated to this momentum exchange, $b=1/\mathcal{Q}$, represents the localization of the valence quarks in the transverse plane, 
relative to the mesons motion as pictorically shown in Fig.~\ref{fig:pion_picture}. If partons are separated more than $b$ that 
particular state will not contribute to the amplitude. Three particle states, e.g. with an extra gluon, will be suppressed by extra 
$1/\mathcal{Q}$ factors, since the probability of finding more than the minimum number of particles bunched up in $1/\mathcal{Q}$ 
decreases as $\mathcal{Q}$ grows. For a simple hard gluon exchange between two fermionic currents the classical dimension 
of the hard scattering amplitude is (mass$)^{-2}$ and since the dependence has to come from external momenta the approximate 
form of the hard amplitude can be expressed as 
\bea
\frac{1}{(xy\,\mathcal{Q}^2)}+\frac{1}{((1-x)(1-y)\,\mathcal{Q}^2)}. \nn
\eea
The end--point values, $x,y\simeq 0,1$, can be problematic as they violate the localization assumption and generate unphysical 
singularities. Indeed, for these values the hard scattering function spreads out in transverse space and it will not be anymore 
concentrated around the $1/\mathcal{Q}$ region. The physical picture corresponds to the case of a fast parton--slow parton couple. 
The slow parton will have $E\approx \Lambda_{QCD}$ and its superposition with more complicated external states is not evidently 
suppressed and indicates a failure of the localization assumption. A very asymmetric, somewhat long--range configuration has to 
be superimposed with the soft external states to estimate the contribution at the end--points. It can be shown, however, that in 
the large momentum exchange limit \cite{Lepage:1980fj,PhysRevLett.24.181,Duncan:1979ny,Braun:1997kw} these contributions are 
typically suppressed by extra factors of $m_q/\mathcal{Q}$ and can be safely neglected as a first order approximation.

%
\subsection{Distribution Amplitudes}
%

\label{sec:B_Ltech}

\begin{figure}[!t]
\includegraphics[scale=0.5]{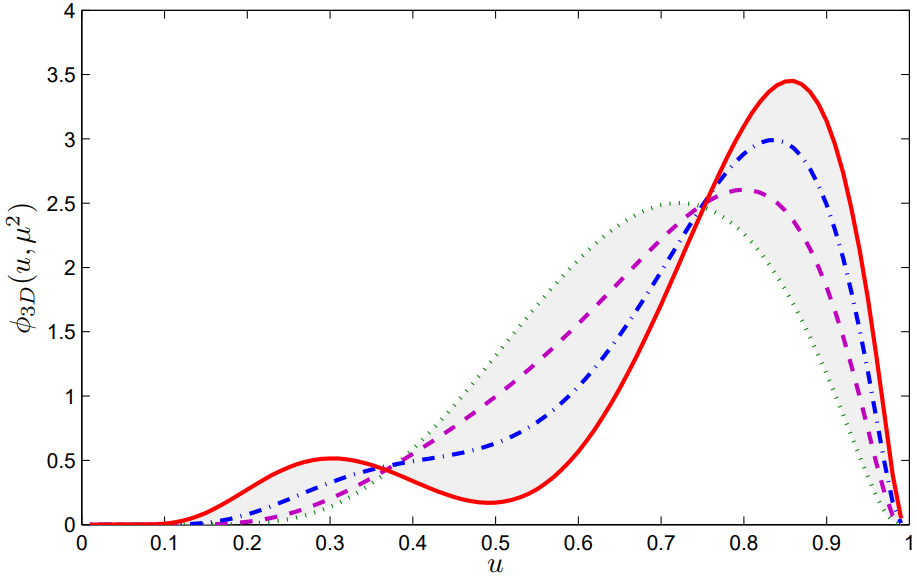}
\caption{The D meson distribution amplitude $\phi_D$ at $\mu=1$ GeV (from \cite{Wu:2013lga}). The dotted, the dashed, the dash-dot 
and the solid lines are for different values of the Gegenbauer momentum $B_D=0, 0.20, 0.40$ and $0.60$ respectively.}
\end{figure}

The simplest way to implement the factorization mechanism described in the previous section is via the theory of QCD exclusive 
processes, firstly developed by Brodsky and Lepage \cite{Lepage:1980fj,Brodsky:1981rp}. The calculation has two main ingredients, 
the momentum DA, $\phi_i$, introduced in Eq.~(\ref{eq:amplitude_fact}) and the hard scattering amplitude. Indeed, the localization 
assumption and the relativistic view discussed at the beginning of Sec.~\ref{sec:hadronization} can be used to build the 
probability distributions of the valence quarks in momentum space, thus recovering their DAs \cite{Radyushkin:1977gp,Huang:2013yya,Wu:2013lga,
Brodsky:1981rp,Lepage:1980fj,Chernyak:1981zz}. 

Let's consider for the moment the case of a light meson $M_L$ of total momentum $P_M$ and let's label with $(\mathbf{p_T},x P_M),$ 
the three--momenta of one of its partons. The 2--component vector $\mathbf{p_T}=(p_x,p_y)$ spans the plane transverse to the direction 
of motion of the meson, while $x$ is the fraction of the meson momentum carried by the considered parton. Imposing the bound--state 
valence quarks approximation (i.e. localization) amounts to taking the part of the probability distribution dependent on its 
transverse momentum as an harmonic oscillator solution, namely:
\beq
\psi_L(x,\mathbf{p}_T) = A_L G_L(x)\exp\left(-\frac{\mathbf{p}^2_T+m_q^2}{8\beta_L^2(1-x)x}\right),
\label{eq:pion_distribution}
\eeq
where $A_L$ is an normalization constant and $\beta_L$ is a mass scale regulating the spread in the transverse plane of $\psi_L$.   
The explicit form of the $\mathbf{p}_T$ dependent part follows from the assumption that in the transverse plane the valence 
quarks lay in a s-wave state. Such a claim is supported by the fact that one is assuming the two partons to be close, thus 
limiting the contributions from states with higher angular momentum. The $x$-dependent function $G_L(x)$ is described by a 
Gegenbauer polynomials expansion~\cite{functions:1953}:
\beq
G_L(x)=\left(1+ B^{(2)}_L C^{3/2}_2(2x-1)+\dots \right).
\eeq
%
The parameters $B^{(n)}_L$ regulate the longitudinal momentum distribution among the partons. By integrating over the 
transverse momentum $\mathbf{p}_T$ down to the scale $\mu$, i.e. up to distances $\sim 1/\mathcal{Q}$, one obtains the 
following expression for the DA function:
\beq
\begin{split}
\phi_L(x,\mu)\propto \beta_L A_L &\sqrt{x(1-x)}G_L(x,\mu) \\ \times & 
      \left(\mathrm{Erf}\left[\sqrt{\frac{m_q^2+\mu^2}{8\beta_L^2x(1-x)}}\right] - 
            \mathrm{Erf}\left[\sqrt{\frac{m_q^2}{8\beta_L^2x(1-x)}}\right]\right).
\end{split}
\eeq
where now the Gegenbauer polynomials get multiplied by scale dependent momenta $B^{(n)}_L(\mu)$, see for example 
Ref.~\cite{PETERLEPAGE1979359,Efremov:1979qk,Mikhailov:1991pt}.
The resulting DA, $\phi_{L}(x,\mu)$ describing the light meson's quark momenta distribution is typically approximated by the 
simmetric function \cite{Lepage:1980fj,Brodsky:1981rp,Szczepaniak:1990dt}:
\bea
 \phi\!_L\!(x) \propto  x(1-x) \,.
 \label{eq:WfunctionL}
\eea

Having worked through the computation for a light meson, it is instructive to consider also the case of a heavy one. Here, 
one has to substitute the argument of the exponential in Eq.(\ref{eq:pion_distribution}) accordingly:
\beq
\frac{\mathbf{p}^2_T+m_q^2}{8\beta_L^2(1-x)x}\to \frac{\mathbf{p}^2_T+m_q^2}{8\beta_H^2(1-x)}+\frac{\mathbf{p}^2_T+m_Q^2}{8\beta_H^2x},
\eeq
where $q$ and $Q$ are the light and heavy partons in the meson. The resulting DA, $\phi_{H}(x,\mu)$, describing the heavy meson's 
quark momenta distribution can be approximated by \cite{Szczepaniak:1990dt}:
\bea
\phi\!_H\!(x) \propto  \left[\frac{\xi^2}{1-x}+\frac{1}{x}-1\right]^{-2}\, ,
\label{eq:WfunctionH}
\eea
where $\xi$ is a small parameter of $O(m_q/m_Q)$, measuring the light/heavy parton asymmetry in the momentum distribution.
The DA functions in Eqs.~(\ref{eq:WfunctionL}) and (\ref{eq:WfunctionH}) are assumed as an ansatz for describing light and heavy 
meson \cite{EFREMOV1980245,Lepage:1980fj,MUELLER1981237} momentum distributions\footnote{A detailed discussion on how adapt this 
DA description to the Kaon sector can be found in \cite{Guerrera:2021yss}.}. To simplify analytical expressions, it is often useful 
to consider the ``very heavy'' meson limit \cite{Bhattacharya:2018msv} by defining the parton masses $m_q=\xi M_I$, $m_Q=(1-\xi)M_I$ 
and assuming the simplified expression 
\bea
\phi_{H}(x) \approx \delta (1-x -\xi) 
\label{veryheavy}
\eea
for the DA function.

Finally one associates to mesons a spinorial representation through the Bethe-Salpeter wave function, $\Psi(x)$, that
carries the quantum numbers of the resonance~\cite{Lepage:1980fj,Szczepaniak:1990dt,Aditya:2012ay,Hazard:2016fnc}. Therefore, 
for pseudoscalar and vector meson one defines respectively:
\bea
\Psi_{\! \mathcal{P}} (x) &=& \frac{\phi(x)}{4}\gamma^5(\slashed{P}_\mathcal{P} + g_\mathcal{P}(x)\,M_\mathcal{P}), \label{eq:wave_PS} \\
\Psi_{\! \mathcal{V}}(x) &=& \frac{\phi(x)}{4}(\sigma^{\alpha\beta}P_{\mathcal{V}\beta}-i g_\mathcal{V}(x) \,M_\mathcal{V}\gamma^\alpha)\epsilon_\alpha(P_\mathcal{V}), \label{eq:wave_V}
\eea
The mass functions $g_{\mathcal{P,V}}(x)$ introduced in Eqs.~(\ref{eq:wave_PS}) and (\ref{eq:wave_V}), are typically assumed to be constant 
and respectively $g_H \approx 1$ and $g_L \ll 1$ for heavy or light mesons. 

With all the ingredients in hand, the hadronic S-matrix elements describing meson decays in ALP can be calculated.



\section{Mesonic Decay amplitudes}
\label{Sec:mesonic}

In this section, the amplitudes for the hadronic and leptonic meson decays into an invisible ALP are derived. The calculation 
of the tree-level amplitudes is performed through the Brodsky--Lepage method, while the hadronization of penguin contributions will 
make use of form factors calculated via LQCD methods.

\subsection{Factorization for $s$--channel processes}
\label{sec:char_tree_level}
\begin{figure}
\centering
\includegraphics[scale=0.32]{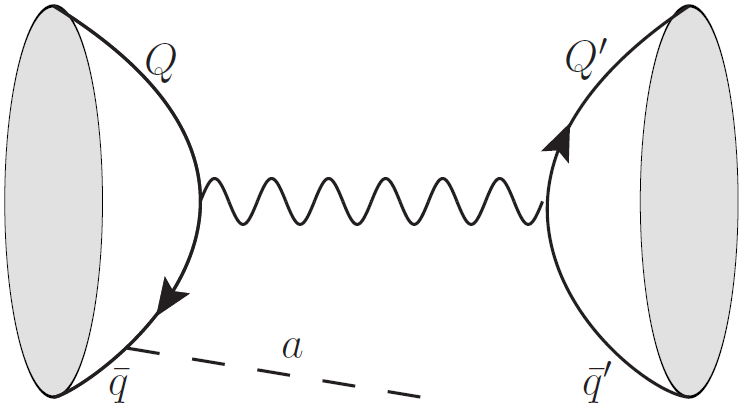}\hspace{.75 cm}\includegraphics[scale=0.32]{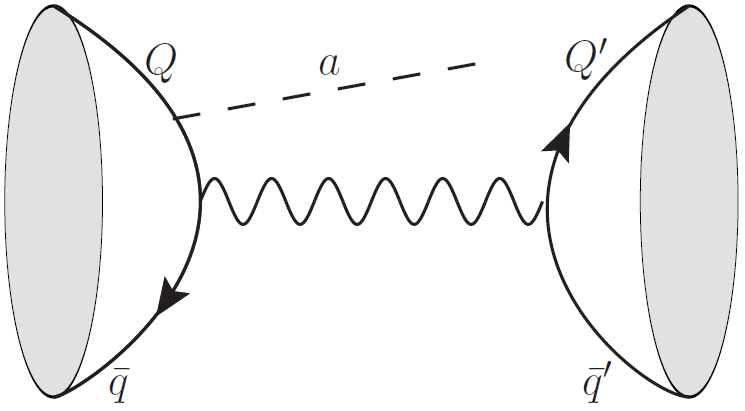}
\caption{Tree-level s-channel of a charged $(\bar{q}Q)$--meson decaying into a charged $(\bar{Q}^\prime q^\prime)$ meson and an ALP. 
Diagrams where the ALP is emitted from the final state meson can be easily obtained.\label{fig:tree_s}}
\end{figure}
A typical s--channel hadronic meson decay process in ALP is shown in Fig.~\ref{fig:tree_s}. Looking at the picture the factorization 
naturally emerges: the amplitude is a product of two uncorrelated vector currents obtained by cutting the diagram along the weak boson 
leg connecting the hadronic external states. Let the initial mesonic state be constituted by the $\bar q Q$ quark pair and the final 
one by the $\bar Q^\prime q^\prime$ quarks. The resulting amplitude will be of the form
\beq
\bra{M_F}(\bar{Q}^\prime\Gamma^{(\mathrm{F})\mu} q^\prime )\ket{0}\bra{0}(\bar q \,\Gamma^{(\mathrm{I})}_\mu Q)\ket{M_I} \label{eq:decomposition},
\eeq
where $M_I$ and $M_F$ are the initial and final mesons and the index $I$ $(F)$ indicates that the ALP is emitted from the initial 
(final) meson quarks. The hadronic to vacuum matrix element is defined as
\beq
\begin{split}
\bra{0} & (\bar q \,\Gamma_\mu Q)\ket{M} = i f_M \int dx \Tr [\Psi_M (x) \Gamma_\mu]\,.
\end{split}
\label{eq:mesonic_hadronization_s}
\eeq
Note that Eqs.~(\ref{eq:wave_PS}), (\ref{eq:wave_V}) and (\ref{eq:mesonic_hadronization_s}) follow a slightly different 
notation with respect to the referred literature. In particular the functions $\phi(x)$ have been normalized to one, in such a way 
that in Eq.~(\ref{eq:mesonic_hadronization_s}) the mesonic decay constants can be explicitly factorized. Moreover, the color 
structure, being trivial in all our processes, has been already explicitly traced. If one of the two operator $\Gamma^{(\mathrm{I,F})}$ 
in Eq.~(\ref{eq:decomposition}) is either $\gamma^\mu\gamma^5$ or $\gamma^\mu$, the hadronization procedure given by 
Eq.~(\ref{eq:mesonic_hadronization_s}) reproduces the usual definitions
\bea
\bra{0}\bar{q}\,\gamma^\mu \,\gamma_5 \,Q \ket{\PS(P_\PS)} & = & i f_\PS P_\PS^\mu \,\label{eq:Mformfactors1} ,\\
\bra{0}\bar{q}\,\gamma^\mu \,Q \ket{\VE(P_\VE)} & = & f_\VE M_\VE\epsilon^\mu(P_\VE) \,,\label{eq:Mformfactors}
\eea
for pseudoscalar and vector mesons, respectively. Thanks to the decorrelation between final and initial state one can obtain 
leptonic and radiative decay contributions by simply replacing one of the hadronic currents in Eq.~(\ref{eq:decomposition}) with 
a leptonic one.

The full amplitude for an s-channel W-mediated hadronic tree level meson decay can be written as: 
\beq
\bra{M_F}(\bar{Q}^\prime\gamma^\mu P_L q^\prime )\ket{0}\bra{0}(\bar q \,\Gamma^{(\mathrm{I})}_\mu Q)\ket{M_I} + 
\bra{M_F}(\bar{Q}^\prime \Gamma^{(\mathrm{F})}_\mu q^\prime )\ket{0}\bra{0}(\bar q \gamma^\mu P_LQ)\ket{M_I}.
\label{eq:def_str}
\eeq
Note that the diagram where the ALP is emitted from the $W$ internal line automatically vanishes, being 
the $W^+W^-$--ALP coupling proportional to the fully antisymmetric 4D tensor. The initial and finale Dirac structures, 
$\Gamma_\mu^{(\mathrm{I,F})}$, can be extracted from the corresponding Feynman diagrams In Fig.~\ref{fig:tree_s} only the case of initial 
meson ALP emission is shown explicitly, with the final meson emission case obtainable straightforwardly. The initial and final 
hard-scattering amplitudes read respectively:
\beq
\begin{split}
\Gamma^{(\mathrm{I})}_\mu & =\frac{4 G_F}{\sqrt{2} f_a}V_{CKM} 
      \left(c_q m_q \frac{\gamma^5\slashed{k}\gamma_\mu P_L}{m_a^2-2 k\cdot P_q}- 
            c_Q m_Q \frac{\gamma_\mu P_L\slashed{k}\gamma^5}{m_a^2-2 k\cdot P_Q} \right) \\
\Gamma^{(\mathrm{F})}_\mu & =\frac{4 G_F}{\sqrt{2} f_a}V_{CKM} 
       \left(c_{Q^\prime} m_{Q^\prime}\frac{\gamma^5\slashed{k}\gamma_\mu P_L}{m_a^2+2 k\cdot P_{Q^\prime}}-
             c_{q^\prime} m_{q^\prime} \frac{\gamma_\mu P_L\slashed{k}\gamma^5}{m_a^2+2 k\cdot P_{q^\prime}} \right), 
\end{split}\label{eq:mesongammas}
\eeq
with $k^\mu$ the ALP 4--momentum. For example, the amplitude for the pseudoscalar-to-pseudoscalar decay, with the ALP emitted 
from the initial meson, arising from the second term in Eq.~(\ref{eq:def_str}), reads: 
\beq
\bra{\PS_F}(\bar{Q}^\prime\gamma^\mu P_L q^\prime )\ket{0}\bra{0} (\bar q \,\Gamma^{(\mathrm{I})}_\mu Q)\ket{\PS_I} =  
i\frac{f_{\PS_F}}{2}P_F^\mu\left(i f_{\PS_I} \int dx \Tr [\Psi_{\PS_I} (x) \Gamma^{(\mathrm{I})}_\mu]\right). 
\label{eq:example_hadronization}
\eeq
The case of vector mesons can be straightforwardly obtained by using Eqs.~(\ref{eq:wave_V}) 
and (\ref{eq:Mformfactors}) instead of Eqs.~(\ref{eq:wave_PS}) and (\ref{eq:Mformfactors1}) in the corresponding amplitudes. In the 
following subsections the explicit expression for the various hadronic and leptonic s-channels decays are presented.

\subsubsection{Mesonic decays}
\label{sec:meson_form}

Using Eqs.~(\ref{eq:mesonic_hadronization_s}), (\ref{eq:mesongammas}) and (\ref{eq:example_hadronization}) and defining
 $P_q=(1-x)P_I$, $P_Q=xP_I$, $P_{q^\prime}=(1-y)P_F$ and $P_{Q\prime}=y P_F$, the s--channel amplitudes 
for a pseudoscalar-to-pseudoscalar meson decay, exemplified by the $B\to K a$ decay, with the ALP radiated from the initial (I) 
or final (F) meson read:
\bea
\mathcal{A}^{(s)}_\mathrm{I} & = & \frac{G_F V_{CKM} f_I f_F(k\cdot P_F)}{\sqrt{2}f_a} M_I \nn \\
  & & \hspace{0.5cm} \int_0^1\! dx \,g_I(x) \left[\frac{c_qm_q\,  \theta(1-x-\delta^M_a)}{m_a^2-2k\cdot P_I(1-x)}-
      \frac{c_Qm_Q\,\theta(x-\delta^M_a)}{m_a^2-2k\cdot P_Ix}\right]\phi_I(x)
\label{eq:a_s_ISR} \\
  \nn \\ \nn \\
\mathcal{A}^{(s)}_\mathrm{F} & = & \frac{G_F V_{CKM} f_I f_F(k\cdot P_I)}{\sqrt{2}f_a} M_F \nn \\ 
  & & \hspace{0.5cm} \int_0^1\! dy\,g_F(y)\left[\frac{c_{Q'}m_{Q'}}{m_a^2+2k\cdot P_Fy}-
      \frac{c_{q'}m_{q'}}{m_a^2+2k\cdot P_F(1-y)}\right]\phi_F(y).
\label{eq:a_s_FSR}
\eea
In the integrals of Eq.~(\ref{eq:a_s_ISR}) an explicit cut-off, $\delta^M_a = m_a/(2M_I)$, has to be introduced for $m_a\neq 0$ to 
remove the unphysical singularities. A simplified analytical expression of the amplitudes of Eq.~(\ref{eq:a_s_ISR}--\ref{eq:a_s_FSR}) 
can be obtained by taking $m_a =0$ and considering the ``very heavy'' meson limit defined in Eq.~(\ref{veryheavy}):
\beq
\left|\mathcal{A}^{(s)}_\mathrm{I} \right|\!\approx\!\frac{G_F V_\mathrm{CKM}f_If_F}{2\sqrt{2}f_a}M_I^2g_H(c_q-c_Q)
\label{eq:happrox_1}
\eeq
\beq
\left|\mathcal{A}^{(s)}_\mathrm{F} \right|\!\approx\!\frac{G_F V_\mathrm{CKM}f_If_F}{2\sqrt{2}f_a}M_F^2g_H(c_{Q^\prime}-c_{q^\prime}),
\label{eq:happrox_2}
\eeq
where $g_H$ is assumed to be constant. This approximation clearly shows the $M^2_{I(F)}$ dependence of the ISR (FSR) amplitude. Therefore, 
in typical pseudoscalar-to-pseudoscalar meson decays the ALP is predominantly emitted form the initial meson. Moreover, as pointed out 
in~\cite{Guerrera:2021yss}, a parametric cancellation occurs in the universal ALP-fermion coupling scenario. This cancellation is still 
partially at work even when the full $\phi(x)$ is used and indicates a possible underestimation of the amplitudes when $c_{q (q')} = 
c_{Q (Q')}$ is chosen.

The s--channel amplitudes for a pseudoscalar-to-vector meson decay, exemplified by the $B\to K^* a$ decay, with the ALP 
radiated from the initial (I) and final (F) meson read:
\bea
\mathcal{B}^{(s)}_\mathrm{I} &= & i \frac{G_F V_{CKM} f_I f_F (k\cdot\epsilon(P_F))}{\sqrt{2}f_a} M_I M_F \nn \\ 
 & & \hspace{0.5cm} \int_0^1 \!dx \,g_I(x)\left[\frac{c_qm_q\, \theta(1-x-\delta^M_a)}{m_a^2-2k\cdot P_I(1-x)}-
  \frac{c_Qm_Q\,\theta(x-\delta^M_a)}{m_a^2-2k\cdot P_Ix}\right]\phi_I(x)
 \label{eq:b_s_ISR} \\
  \nn \\ \nn \\
\mathcal{B}^{(s)}_\mathrm{F} & = &i \frac{G_F V_{CKM} f_I f_F}{\sqrt{2}f_a} \epsilon^\alpha(P_F) P_F^\beta(k^\beta P_I^\alpha- 
  k^\alpha P_I^\beta)  \nn \\ 
 & & \hspace{0.5cm} \int_0^1 \!dy \left[\frac{c_{Q'}m_{Q'}}{m_a^2+2k\cdot P_Fy}\, - \, 
  \frac{c_{q'}m_{q'}}{m_a^2+2k\cdot P_F(1-y)}\right]\phi_F(y),
\label{eq:b_s_FSR}
\eea
where $\epsilon^\alpha$ is the polarization of the vector resonance. The results for vector-to-pseudoscalar meson decay read:
\bea
\mathcal{C}^{(s)}_\mathrm{I} 
&=& -i\frac{G_F V_{CKM} f_If_F}{\sqrt{2}f_a} \epsilon^\alpha(P_I)P_I^\beta (k^\alpha P_F^\beta- 
  k^\beta P_F^\alpha) \nn \\ 
& & \hspace{0.5cm} \int_0^1 \!dx \left[\frac{c_q m_q\, \theta(1-x-\delta^M_a)}{m_a^2-2k\cdot P_I(1-x)} - 
  \frac{c_Q m_Q\, \theta(x-\delta^M_a)}{m_a^2-2k\cdot P_Ix}\right]\phi_I(x). \label{eq:c_s_ISR} \\
  \nn \\ \nn \\
\mathcal{C}^{(s)}_\mathrm{F} 
&=& -i\frac{G_F V_{CKM}f_If_F(k\cdot\epsilon(P_F)) }{\sqrt{2}f_a} M_I M_F \nn \\ 
& & \hspace{0.5cm} \int_0^1 \! dy \, g_F(y) \left[\frac{c_{Q'}m_{Q'}}{m_a^2+2k\cdot P_Fy}-
    \frac{c_{q'}m_{q'}}{m_a^2+2k\cdot P_F(1-y)}\right]\,\phi_F(y).
 \label{eq:c_s_FSR}
\eea
%
One can easily show that also in these cases, for $m_a=0$ and assuming the ``very heavy'' meson limit, simple expressions for the 
amplitudes of the $\mathcal{B}$ and $\mathcal{C}$--type decays can be recovered, exhibiting a meson $M^2_{I(F)}$ dependence and 
a parametric cancellation for a universal ALP-fermion coupling, similarly to the results of Eqs.~(\ref{eq:happrox_1}--\ref{eq:happrox_2}).

Finally the amplitudes for vector-to-vector meson decays read: 
%
\bea
\mathcal{D}^{(s)}_\mathrm{I} &=& -i\frac{G_F V_{CKM} f_I f_F}{\sqrt{2}f_a} \epsilon^\alpha(P_I) \epsilon^\mu(P_F)  P_I^\beta M_F  \nn \\ 
& & \hspace{0.5cm} \int_0^1 \!dx \left[\frac{c_q m_q\,\theta(1-x-\delta^M_a)}{m_a^2-2k\cdot P_I(1-x)}(\varepsilon_{\alpha\beta\mu\rho}k^\rho 
    + i k^\beta g^{\alpha\mu} -i k^\alpha g^{\beta\mu}) \, +\, \right. \nn \\ 
& & \hspace{1.5cm} \,\,\,\,+ \left. \frac{c_Q m_Q\,\theta(x-\delta^M_a)}{m_a^2-2k\cdot P_Ix}(\varepsilon_{\alpha\beta\mu\rho}k^\rho + i k^\alpha g^{\beta\mu} 
    - i k^\beta g^{\alpha\mu} )\right]\phi_I(x) \label{eq:d_s_ISR} \\
\nn \\ \nn \\
\mathcal{D}^{(s)}_\mathrm{F} &=& -i\frac{G_F V_{CKM} f_I f_F}{\sqrt{2}f_a} \epsilon^\alpha(P_F) \epsilon^\mu(P_I) P_F^\beta M_I  \nn \\ 
& & \hspace{0.2cm} \int_0^1 \!dy \left[\frac{c_{Q'} m_{Q'}}{m_a^2+2k\cdot P_Fy}(\varepsilon_{\alpha\beta\mu\rho}k^\rho 
   + i k^\beta g^{\alpha\mu} -i k^\alpha g^{\beta\mu}) \, +\, \right. \nn \\
& &  \hspace{1.2cm} \,\,\,+ \left.\frac{c_{q'} m_{q'}}{m_a^2+2k\cdot P_F(1-y)}(\varepsilon_{\alpha\beta\mu\rho}k^\rho 
   + i k^\alpha g^{\beta\mu} -i k^\beta g^{\alpha\mu} )\right]\!\phi_F(y). \label{eq:d_s_FSR}
\eea
In the ``very heavy'' meson limit the vector-to-vector amplitudes read:
\bea
\left|\mathcal{D}^{(s)}_\mathrm{I}\right| &\approx & \frac{f_I f_F G_F M_I^2 V_\mathrm{CKM}}{4 f_a}
                        \sqrt{(c_Q-c_q)^2+4\frac{M_F^2}{M_I^2}(c_Q^2+c_q^2)}, \label{heavyDI} \\
\left|\mathcal{D}^{(s)}_\mathrm{F}\right| &\approx & \frac{f_I f_F G_F M_F^2 V_\mathrm{CKM}}{4 f_a}
                        \sqrt{(c_Q^\prime-c_q^\prime)^2+4\frac{M_I^2}{M_F^2}(c_Q^{\prime2}+c_q^{\prime2})}. \label{heavyDF}
\eea
The parametric cancellation in this case is not completely at work as the epsilon tensor in 
Eqs.~(\ref{eq:d_s_ISR}-\ref{eq:d_s_FSR}) introduces an extra term proportional to the sum of the couplings squared. Notice 
also that, in the universal ALP-fermion coupling scenario, both the ISR and FSR amplitudes get proportional to the product 
$M_I \, M_F$ and there is no clear suppression of the FSR ALP process with respect to the ISR one, contrary to what happens 
for all the processes described by the $\mathcal{A}, \mathcal{B}$ and $\mathcal{C}$ amplitudes.

\subsubsection{Leptonic decay}
\label{sec:semilept_form}

\begin{figure}[t]
\centering
\includegraphics[scale=0.17]{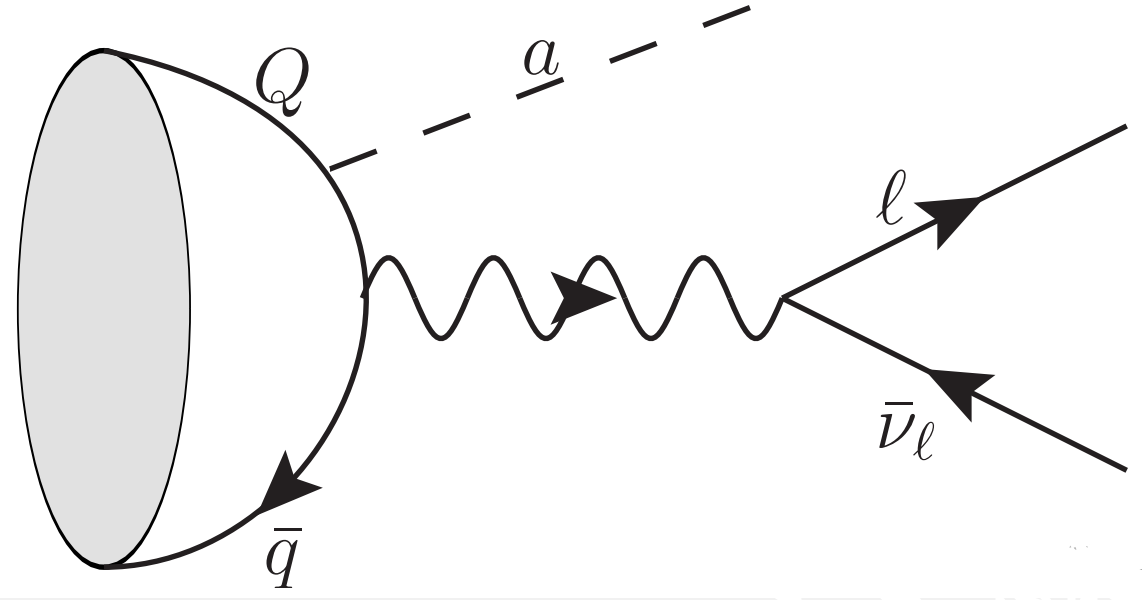}\hspace{1.3 cm}\includegraphics[scale=0.17]{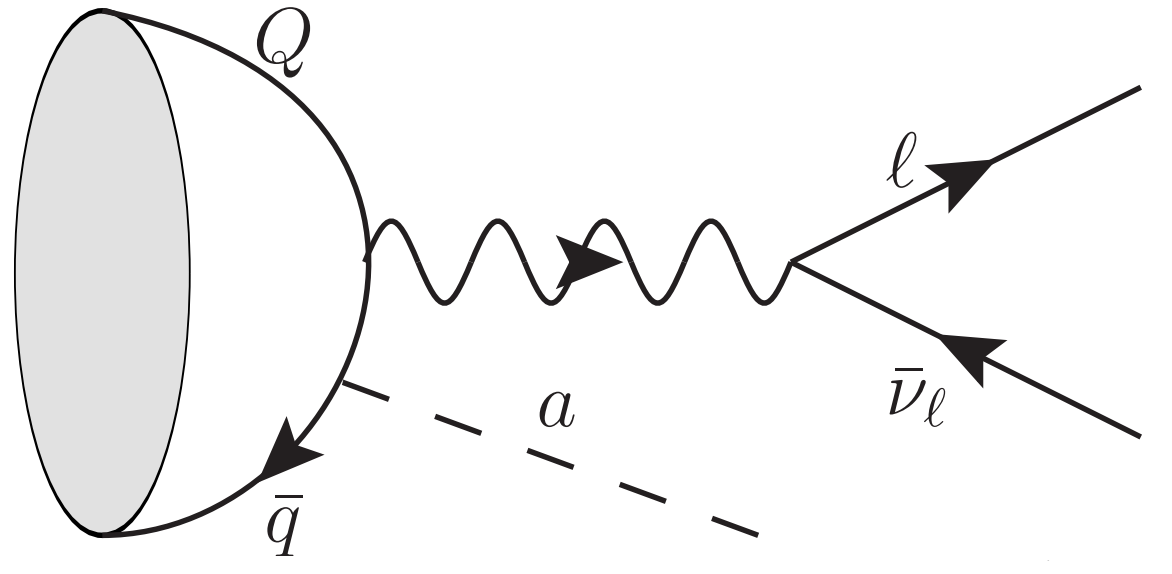}
\caption{Tree level contributions to the $M \to \ell\, \nu_\ell \, a$ amplitude, with the ALP emitted from the $M$ 
meson. The diagram where the ALP is emitted from the charged lepton is straightforward.}
\label{fig:tree_charged_lept}
\end{figure}

For completeness we report here, briefly, the semileptonic meson decays, $M \to \ell\, \nu_\ell a$, derived in details 
in~\cite{Gallo:2021ame}. In Fig.~\ref{fig:tree_charged_lept} the diagrams where the ALP is emitted from the meson are drawn. 
The diagrams where the ALP is emitted by the charged leptons follow straightforwardly. These amplitudes can be factorized as 
\beq
\bra{0}  (\bar q \,\Gamma_\mu^{(h)} Q)\ket{M}(\bar{\ell}\gamma^\mu P_L \nu) + 
\bra{0}  (\bar q \,\gamma^\mu P_L Q)\ket{M}(\bar{\ell}\,\Gamma_\mu^{(\ell)}\, \nu),
\eeq
with $\Gamma_\mu^{(h)}$ and $\Gamma_\mu^{(\ell)}$ the Dirac structures related to the hadronic and leptonic emission, respectively.
Using the methods introduced in Eqs.~(\ref{eq:mesonic_hadronization_s}--\ref{eq:Mformfactors1}) one obtains 
\beq
\begin{split}
\mathcal{E}_h=\frac{4 i G_F V_\mathrm{CKM}}{\sqrt{2}}\frac{f_M}{f_a}\frac{M_M^2}{2k\cdot P_M}\, 
                    \Big[ c_Q \frac{m_Q}{M_M}  \Phi^{(Q)}_M(m_a^2)
                      - c_q \frac{m_q}{M_M} \Phi^{(q)}_M (m_a^2) \Big]
                      \left(\bar{\ell} \, \slashed{k} \,P_L \, \nu_\ell \right).\label{eq:MMALP} 
\end{split}
\eeq
for the amplitudes associated to the hadronic ALP emission. The functions $\Phi^{(q,Q)}_M (m_a^2)$ contain the integrals over 
the quark momentum fraction and are given by:
\bea
\Phi^{(q)}_M (m_a^2) &=& \int^{1-\delta_M}_0 \frac{k \cdot P_M}{m_a^2-2\,(1-x)\,k\cdot P_M} \, \phi_M(x) \,g_M(x)\, dx \nn \\
\Phi^{(Q)}_M (m_a^2) &=& \int^1_{\delta_M} \frac{k \cdot P_M}{m_a^2-2\,x\,k\cdot P_M}     \, \phi_M(x) \,g_M(x)\, dx \,.
\label{eq:leptonic_integrals}
\eea
One can obtain a simple expression for the hadronic amplitude by employing the ``very heavy'' meson approximation discussed in 
Sec.~\ref{sec:meson_form}. For $m_a=0$ one has
\beq
\mathcal{E}_h\approx i G_F V_\mathrm{CKM}\frac{f_M}{\sqrt{2}f_a}\frac{M_M^2}{k\cdot P_M}(c_q-c_Q) \left(\bar{\ell} \, \slashed{k} \,P_L \, \nu_\ell \right).
\eeq
The leptonic decay amplitude for the lepton ALP--emission process can be easily obtained by using the definition of the meson 
form factors of Eq.~(\ref{eq:Mformfactors1}), giving
\bea
\mathcal{E}_\ell &=& - \frac{4\,i\, G_F}{\sqrt{2}} V_{qQ} \,\frac{f_M}{f_a}  \left[  c_\ell\,m_\ell\, 
\left(\bar{\ell} \, P_L \,\nu_\ell \right) - 
\frac{ c_\ell \, m^2_\ell}{m_a^2\,+\, 2\, k\cdot p_\ell} \left(\bar{\ell} \, \slashed{k}\, P_L \,\nu_\ell \right) \right] \,.
\label{eq:MLALP}
\eea
assuming vanishing neutrino masses.

\subsection{Factorization of $t$--channel processes}
\label{sec:neut_tree_level}

\begin{figure}
\centering
\includegraphics[scale=0.33]{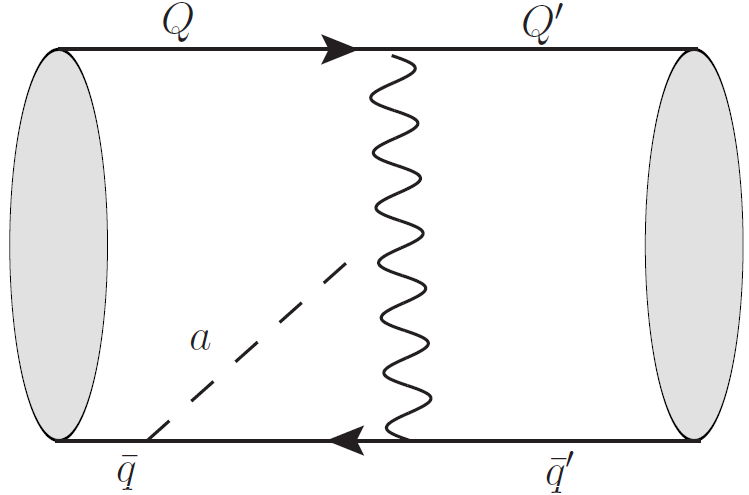}\hspace{.5 cm}\includegraphics[scale=0.33]{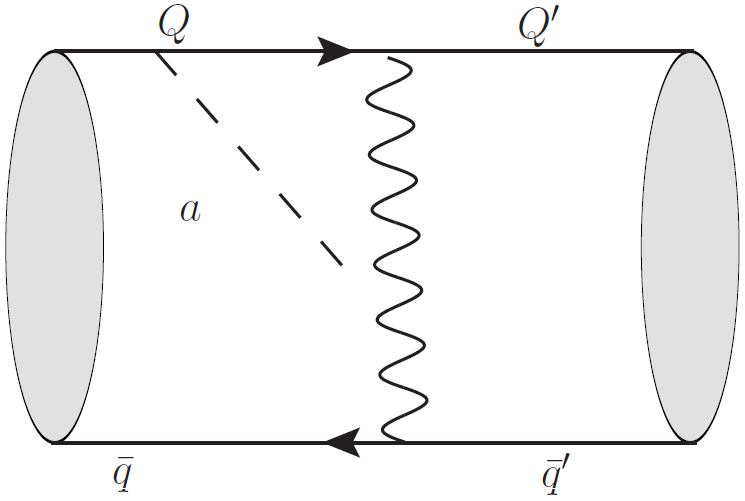}
\caption{Tree-level t-channel of a neutral $(\bar{q}Q)$--meson decaying into a neutral $(\bar{Q}^\prime q^\prime)$ meson and an ALP.
Diagrams where the ALP is emitted from the final state meson can be easily obtained. Similar diagrams can be depicted for the 
CP conjugate process.}\label{fig:fig_t}
\end{figure}

Following a similar approach to the one used in the previous subsection one can now study meson decays occurring through a 
$t$--channel $W$ exchange. In Fig.~\ref{fig:fig_t}, typical diagrams with the ALP emitted from the initial meson state are 
depicted. Once again, diagrams with the ALP emitted from the final state are obvious, while the diagram where the ALP is emitted 
from the $W$ internal line automatically vanishes. In this case the hadronic process can be written:
\beq
\bra{M_F}(\bar Q' \,\Gamma^{(Q)}_\mu \,Q)(\bar{q} \, \gamma^\mu P_L \,q' )\ket{M_I} + 
\bra{M_F}(\bar Q' \,\gamma^\mu P_L \, Q)(\bar{q}\, \Gamma^{(\bar q)}_\mu \,q')\ket{M_I}, \label{eq:t_decomposition}
\eeq
where $\Gamma^{(Q,\bar{q})}$ are the Feynman amplitudes corresponding to the ALP emission from a quark or an antiquark line. Note 
that for processes in the $t$--channel, there are no trivial mesonic currents representing either the initial or final meson, and 
then the full Brodsky--Lepage machinery is always required for calculating these amplitudes.
The tree--level hard scattering of the diagrams depicted in Fig.\ref{fig:fig_t}, gives: 
\bea
\Gamma^{(Q)}_\mu & = & \frac{4 G_F}{\sqrt{2} f_a} V_{CKM} \left(c_{Q'} m_{Q'} \frac{\gamma^5\slashed{k}\gamma_\mu P_L}{m_a^2+2 k\cdot P_{Q'}}
                  - c_Q m_Q \frac{\gamma_\mu P_L\slashed{k}\gamma^5}{m_a^2-2 k\cdot P_Q} \right)\label{eq:gamma_Q} \\
\Gamma^{(\bar q)}_\mu & = & \frac{4 G_F}{\sqrt{2} f_a}V_{CKM}\left(c_{q} m_{q} \frac{\gamma^5\slashed{k}\gamma_\mu P_L}{m_a^2-2 k\cdot P_{q}}
                  - c_{q'} m_{q'} \frac{\gamma_\mu P_L\slashed{k}\gamma^5}{m_a^2+2 k\cdot P_{q'}} \right). \label{eq:gamma_qbar}
\eea
Using the procedure described in Ref.~\cite{Szczepaniak:1990dt}, one obtains, for the $t$--channel factorization:
\bea
\bra{M_F}(\bar Q'\, \Gamma^{(Q)}_\mu \,Q)(\bar{q} \, \gamma^{\mu}P_L\, q' )\ket{M_I}\!\! &=&\!  - 
  \frac{f_{M_F}f_{M_I}}{\sqrt{2}} \! \int \! dx\, dy \,\Tr [ \Psi_{M_I}(x)\,\gamma^{\mu}P_L\,\Psi_{M_F}(y)\,\Gamma^{(Q)}_\mu]\,,\nn\\
\bra{M_F}(\bar Q'\, \gamma^{\mu}P_L\ \,Q)(\bar{q} \, \Gamma^{(\bar q)}_\mu\, q' )\ket{M_I}\!\! &=&\!  - 
  \frac{f_{M_F}f_{M_I}}{\sqrt{2}} \! \int \! dx\, dy \,\Tr [ \Psi_{M_I}(x)\,\Gamma^{(\bar{q})}_\mu\,\Psi_{M_F}(y)\,\gamma^{\mu}P_L]\,.\nn\\ \label{eq:mesonic_hadronization}
\eea 
Substituting $\Gamma^{(Q,\bar{q})}_\mu$ according to Eqs.~(\ref{eq:gamma_Q}--\ref{eq:gamma_qbar}) one finds the hadronized amplitudes. 
It is then phenomenologically convenient to separate ISR and FSR amplitudes.
The pseudoscalar-to-pseudoscalar meson decay amplitudes read:
\bea 
\mathcal{A}^{(t)}_\mathrm{I} & = & \frac{G_F V_{CKM} f_I f_F (k\cdot P_F) }{2f_a} M_I \nn \\ 
 & & \hspace{0.5cm} \int_0^1 dx \, g_I(x)\left[\frac{c_Qm_Q\,\theta(x-\delta^M_a)}{m_a^2-2k\cdot P_Ix}-\frac{c_qm_q\,\theta(1-x-\delta^M_a)}{m_a^2-2k\cdot P_I(1-x)}\right]\phi_I(x)\, , 
 \label{eq:a_t_ISR} \\
   \nn \\ \nn \\
\mathcal{A}^{(t)}_\mathrm{F} & = & \frac{G_F V_{CKM} f_I f_F (k \cdot P_I) }{2f_a} M_F \nn \\ 
 & & \hspace{0.5cm} \int_0^1 dy \,g_F(y) \left[\frac{c_{q'}m_{q'}}{m_a^2+2k\cdot P_F(1-y)}-\frac{c_{Q'}m_{Q'}}{m_a^2+2k\cdot P_Fy}\right]\phi_F(y) \, , 
 \label{eq:a_t_FSR}
\eea
while the pseudoscalar-to-vector $t$--channel transitions read:
\bea
\mathcal{B}^{(t)}_\mathrm{I} &=& i\frac{G_F V_{CKM} f_I f_F(k\cdot\epsilon(P_F)) }{2f_a} M_I M_F \nn \\ 
& & \hspace{0.5cm} \int_0^1 dx  \, g_I(x) \left[\frac{c_{Q}m_{Q}\,\theta(x-\delta^M_a)}{m_a^2-2k\cdot P_I x}-\frac{c_{q}m_{q}\,\theta(1-x-\delta^M_a)}{m_a^2-2k\cdot P_I(1-x)}\right]\phi_F(x)\, , 
\label{eq:b_t_ISR} \\
  \nn \\ \nn \\
\mathcal{B}^{(t)}_\mathrm{F} &=& i\frac{G_F V_{CKM} f_I f_F}{2f_a} \epsilon^\alpha(P_F)P_F^\beta(k^\beta P_I^\alpha-k^\alpha P_I^\beta)  \nn \\ 
& & \hspace{0.5cm} \int_0^1 dy\Big[\frac{c_{Q'}m_{Q'}}{m_a^2+2k\cdot P_Fy} - \frac{c_{q'}m_{q'}}{m_a^2+2k\cdot P_F(1-y)}\Big] \phi_F(y)\, .
\label{eq:b_t_FSR}
\eea
Finally, the vector-to-pseudoscalar and vector-to-vector decays amplitudes are given by:
\bea
\mathcal{C}^{(t)}_\mathrm{I} &=& -i\frac{G_F V_{CKM} f_I f_F}{2f_a}  \epsilon^\alpha(P_I)P_I^\beta(k^\beta P_F^\alpha - k^\alpha P_F^\beta)   \nn \\ 
& & \hspace{0.5cm} \int_0^1 dx \Big[\frac{c_Q m_Q\,\theta(x-\delta^M_a)}{m_a^2-2k\cdot P_Ix} - \frac{c_q m_q\,\theta(1-x-\delta^M_a)}{m_a^2-2k\cdot P_I(1-x)}
\Big] \phi_I(x).\label{eq:c_t_ISR} \\
   \nn \\ \nn \\
\mathcal{C}^{(t)}_\mathrm{F} &=&- i\frac{G_FV_{CKM} f_I f_F (k\cdot \epsilon(P_F))}{2f_a} M_I M_F \nn \\ 
& & \hspace{0.5cm} \int_0^1 dy \, g_F(y) \Big[\frac{c_{Q'}m_{Q'}}{m_a^2+2k\cdot P_Fy}-\frac{c_{q'}m_{q'}}{m_a^2+2k\cdot P_F(1-y)}\Big]\phi_F(y)\, ,
\label{eq:c_t_FSR}
\eea
and by
\bea
\mathcal{D}^{(t)}_\mathrm{I} &=& -i\frac{G_F V_{CKM} f_I f_F}{2f_a} \epsilon^\alpha(P_I)\epsilon^\mu(P_F) P_I^\beta M_F  \nn  \\ 
& & \hspace{0.5cm} \int_0^1 dx \Big[\frac{c_Q m_Q\,\theta(x-\delta^M_a)}{m_a^2-2k\cdot P_Ix} 
(\varepsilon_{\alpha\beta\mu\rho} k^\rho - i k^\beta g^{\alpha\mu} + i k^\alpha g^{\beta\mu}) + \nn \\ 
& & \hspace{1.5cm}\,\,\,\, \frac{c_q m_q\,\theta(1-x-\delta^M_a)}{m_a^2-2k\cdot P_I(1-x)}
 (\varepsilon_{\alpha\beta\mu\rho} k^\rho - i k^\alpha g^{\beta\mu} +i k^\beta g^{\alpha\mu})\Big]\phi_I(x).\label{eq:d_t_ISR} \\ 
  \nn \\ \nn \\
\mathcal{D}^{(t)}_\mathrm{F} &=& -i\frac{G_F V_{CKM} f_I f_F}{2f_a}\epsilon^\alpha(P_I)  \epsilon^\mu(P_F)P_F^\beta M_I  \nn \\  
& & \hspace{0.5cm} \int_0^1 dy \Big[\frac{c_{Q'} m_{Q'}}{m_a^2+2k\cdot P_Fy}
 (\varepsilon_{\alpha\beta\mu\rho} k^\rho - i k^\beta g^{\alpha\mu} + i k^\mu g^{\alpha\beta}) + \nn \\ 
& & \hspace{1.5cm}\,\,\,\, \frac{c_{q'} m_{q'}}{m_a^2+2k\cdot P_F(1-y)}
 (\varepsilon_{\alpha\beta\mu\rho} k^\rho - i k^\mu g^{\alpha\beta} +i k^\beta g^{\alpha\mu})\Big] \phi_F(y).\label{eq:d_t_FSR}
\eea
It is interesting to note that the results derived here are very symmetric with the $s$-channel case, up to a sign and a 
factor of $\sqrt{2}$. The only exception is given by the vector-to-vector decays where a minus sign is present in parts of the
tensor structure w.r.t. the $s$-channel case.
One can apply the ``very heavy" approximation used in Sec.~\ref{sec:meson_form} to derive simple analytic results also for 
the $t$-channel amplitudes. Expressions similar to the ones shown in Eqs.~(\ref{eq:happrox_1}--\ref{eq:happrox_2}) and  
Eqs.~(\ref{heavyDI}--\ref{heavyDF}) can be obtained also in the $t$-channel case.

%

\subsection{Penguin Hadronization}
\label{subs:penguin}

\begin{figure}[!th]\center
\includegraphics[scale=0.13]{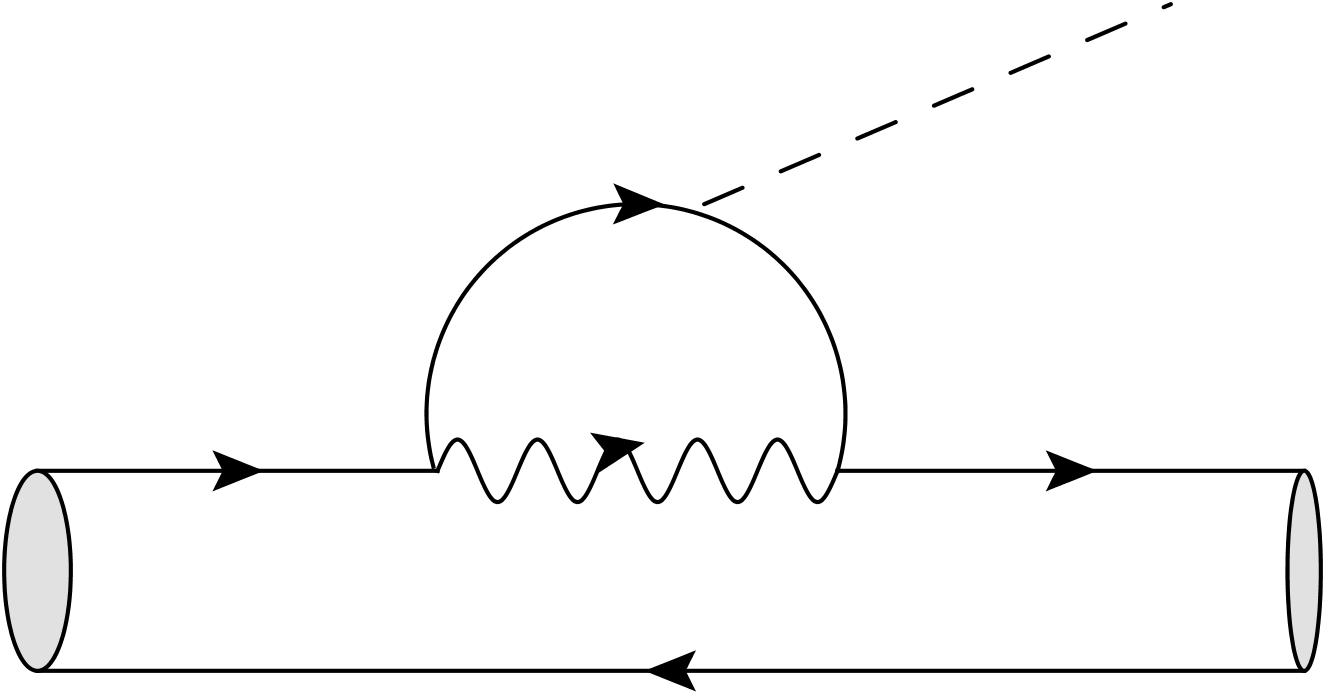}\hspace{1 cm}\includegraphics[scale=0.13]{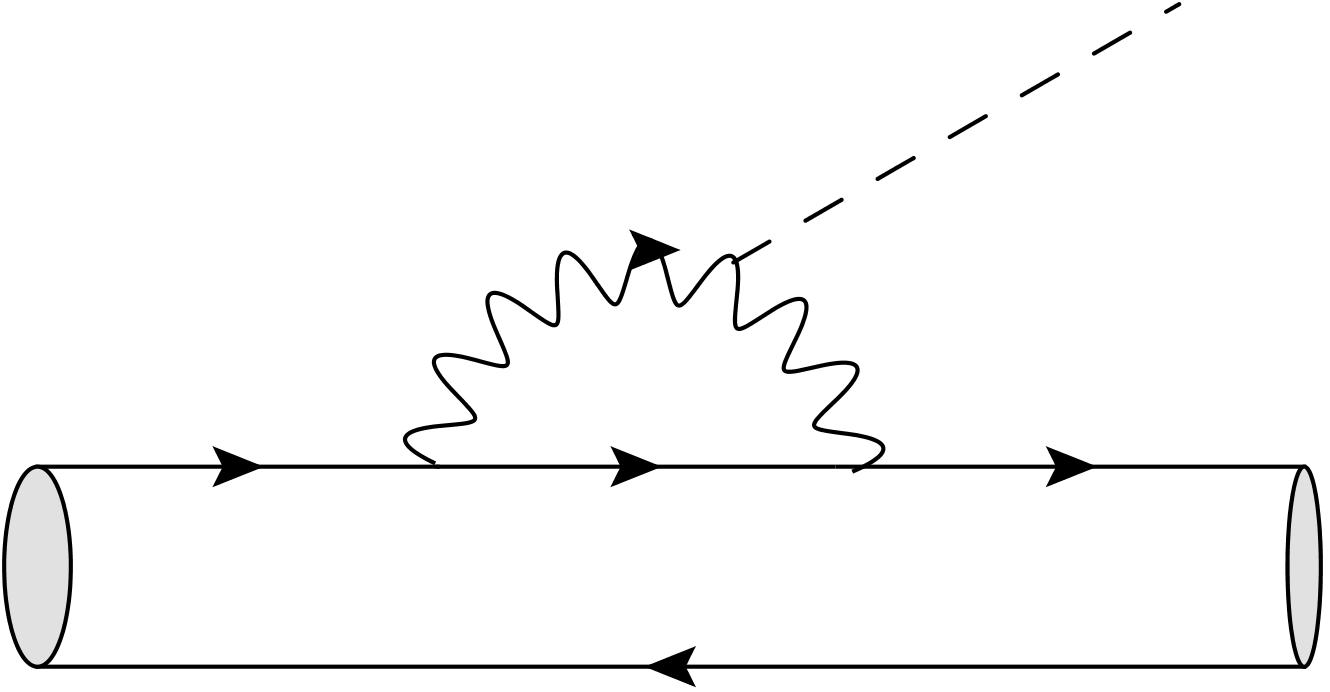}
\caption{Dominant one-loop penguin contributions to meson decay in ALP. \label{fig:figure_loop}}
\end{figure}

Flavor changing neutral current meson decays in ALP will often receive the dominant contributions from the one--loop penguin 
diagrams \cite{Izaguirre:2016dfi,Bauer:2017ris,Gavela:2019wzg}, shown in Fig.~\ref{fig:figure_loop}. 
In this kind of processes, only one quark line participates actively to the ALP emission, with the other quark playing  
the role of spectator. 

The hadronic matrix element for pseudoscalar-to-pseudoscalar meson decays is mediated by a vector current customarily 
factorised as:
\bea
\bra{\PS_F}\bar{q}\gamma^\mu Q\ket{\PS_I}=f_+(k^2)(P_{I}+P_{F})^\mu+f_-(k^2)\,k^\mu
\label{eq:loop_formfactors}
\eea
with $k=P_{I}-P_{F}$. The $f_{+,-}(k^2)$ form factors can be obtained from LQCD calculations \cite{Carrasco:2016kpy,
Gubernari:2018wyi,Ball:1993tp,Ball:2004rg,Ball:2004ye,Wang:2008ci,Wang:2008xt,Issadykov:2015iba,Lubicz:2017syv,
FermilabLattice:2017wmk,Cooper:2020wnj}, and are transition specific. Using Eq.~(\ref{eq:loop_formfactors}), the penguin 
contribution to the $\mathcal{P}_I \to \mathcal{P}_F \,a$ decay amplitude, assuming flavor diagonal quarks-ALP couplings, reads:
\beq
\begin{split}
&{\mathcal G}_{\PS_I\to\PS_F} = \frac{G_F\, m^2_q}{2\sqrt{2}\pi^2}\frac{M^2_I}{f_a}\left(1-\frac{M^2_F}{M^2_I}\right)\!
\left[f_+(k^2) + \frac{k^2}{M^2_I-M^2_F} f_-(k^2)\right]\!\sum_{f}\! \!c^{(f)}_{ij}.
\label{loopAmplitudeK+}        
\end{split}              
\eeq
The coefficient $c^{(f)}_{ij}$ has been opportunely normalized to factorize out the heaviest quark mass running in the loop, 
here dubbed $m_q$, and is defined as:
\beq
c^{(f)}_{ij} = V_{fi}V_{fj}^* \left[3\,c_W \frac{g(x_f)}{x_q}-\frac{c_{f}\,x_f}{4\,x_q} \ln\left(\frac{f_a^2}{m_f^2}\right)\right] 
\qquad \mbox{with} \quad \left(x_f \equiv \frac{m_f^2}{m_W^2}\right)\,.
\label{ALPflavorViolfirst}
\eeq
The penguin contribution where the ALP emitted from the internal $W$ line is included here for completeness, even if in the 
following phenomenological analysis $c_W=0$ will be assumed\footnote{ The contribution due to weak boson-ALP coupling and the 
interplay between quark and weak boson ALP coupling has been considered for example in~\cite{Gavela:2019wzg}.}. 
One-loop diagrams, with the ALP emitted from the initial/final quarks are suppressed by at least an extra $m_{f}^2/m_W^2$ factor 
(being $m_f$ the mass of the initial/final quark) with respect to the penguin contributions, as they arise at third order in 
the external momenta expansion. In the case of the $K$ and $D$ meson they can be safely neglected even compared to the  
tree level contributions. For $B$ mesons, instead, these contributions are roughly of the same order of the tree-level ones. 
Therefore, the complete 1--loop renormalization should be performed to be able to extract information on the ALP-external fermion 
couplings. 

In the case of vector-to-pseudoscalar transitions the hadronic matrix element can be divided in terms of four independent form factors. 
See for example \cite{Ball:2004rg} for the general expression. For the specific type of decays considered here, the only surviving 
form factor is given by:
\beq
\bra{\PS}\bar{q}\gamma^\mu\gamma_5Q\ket{\VE} = 2\,i \,M_\VE A_0(k^2) \frac{\left(\epsilon(P_\VE)\cdot k \right) k^\mu}{k^2}.
\label{eq:v_to_ps_FF}
\eeq
For numerical evaluation the $A_0(k^2)$ results explicitly reported in~\cite{Ball:1993tp,Ball:2004rg,Wang:2008xt,Wang:2008ci,
Fu:2014pba,Bharucha:2015bzk,Issadykov:2015iba,Gubernari:2018wyi,McLean:2019sds,McLean:2019jll} has been used. For strange 
meson decays, when not available, the form factors have be chosen to be 1, assuming an exact $SU(3)$ flavor symmetry. Finally, 
following the conventions of \cite{Bharucha:2015bzk}, one obtains
\beq
\mathcal{G}_{\VE\to\PS} = i \frac{G_F m_q^2}{\sqrt{2}\pi^2f_a} M_\VE A_0 (k^2) 
    \left(\epsilon(P_\VE)\cdot k \right)\,\sum_{f} c^{(f)}_{ij} 
\label{loopAmplitudeVector}    
\eeq
with $m_q$ the mass of the heaviest quark running in the loop and the coefficient $c^{(f)}_{ij}$ defined previously in 
Eq.~(\ref{ALPflavorViolfirst}).
The case of vector-to-pseudoscalar can be obtained from Eq.~(\ref{loopAmplitudeVector}) by exchanging $M_\VE$ 
with $M_\PS$ and taking the conjugate of the amplitude.


\section{Phenomenology of Invisible ALP and Mesons}
\label{sec:invisible_bounds_ff}

From the meson decay amplitudes calculated in the previous section, stringent limits on the ALP-fermion Effective Lagrangian 
of Eq.~(\ref{eq:lag_def_c}) can now be derived. The two different classes of meson decays are going to be discussed separately. 
Hadronic meson decays in ALPs are expected in general, to be the most constraining processes to test ALP--quark couplings 
in the sub-GeV ALP mass range, mainly thanks to the high precision experimental results of the Kaon sector 
\cite{NA62:2020pwi,CortinaGil:2020fcx,CortinaGil:2021nts,CortinaGil:2021gga,Ahn:2018mvc}. Nonetheless, very promising results 
are expected from $B$--factories \cite{Masso:1995tw,Bevan:2014iga,Dolan:2017osp,Kou:2018nap,CidVidal:2018blh,deNiverville:2018hrc,
Belle:2017oht}, for ALP masses up to few GeV \cite{Dolan:2014ska,Izaguirre:2016dfi,Gavela:2019wzg,MartinCamalich:2020dfe}. 
Semileptonic decays, on the other hand are useful to test ALP--lepton couplings producing new bounds in the KeV--GeV range 
for the all the ALP-charged lepton couplings, whereas most of the limits on ALP-quark couplings are not competitive with the 
ones obtained from the hadronic decay channels.

\subsection{Hadronic final states}
\label{sec:true_pheno}
In Sec.~\ref{Sec:mesonic} the tree-level and one-loop (penguin) contributions to the hadronic meson decays into an invisible 
ALP, $M_I\to M_F \, a$, have been derived. In Tab.~\ref{tab:constraints} the amplitudes of several charged meson decays are 
collected. For definiteness, $m_a=0$, $f_a=1$ TeV and $c_f=\pm 1$ have been used. As noticed in Sec.~\ref{sec:char_tree_level} and 
\ref{sec:neut_tree_level}, accidental cancellation can occurs in the tree--level amplitudes, depending on the relative sign between 
$c_{Q^{(\prime )}}$ and $c_{q^{(\prime )}}$ for all the processes but vector-to-vector decays, see Eqs.~(\ref{eq:a_s_ISR}-\ref{eq:c_s_FSR}) 
and Eqs.~(\ref{eq:a_t_ISR}-\ref{eq:c_t_FSR}). To make evident the impact of this accidental cancellation, the tree-level results in 
Tab.~\ref{tab:constraints} has been shown with a $(min-max)$ interval, obtained by setting $c_{Q^{(\prime )}}/c_{q^{(\prime )}}= (+1,-1)$ 
respectively. The origin of this parametric cancellation has been proved analytically by \cite{Guerrera:2021yss} both in the ``very 
light'' and ``very heavy'' meson limit. In Sec.~\ref{sec:char_tree_level} the results for the ``very heavy'' meson limits have 
been explicitly shown for the $\mathcal{A}$-type decays. From the results of Tab.~\ref{tab:constraints}, one learns which is the 
effectiveness of this parametric cancellation, once the numerical integration is performed using the non-approximated heavy 
DA function of Eq.~(\ref{eq:WfunctionH}). Depending on the specific decay channel, the tree-level decay rate 
can change from one to two orders of magnitude. Notice that in Tab.~\ref{tab:constraints} no vector-to-vector decay is 
presented, being still experimentally marginal. 
\begin{table}[t!]
\centering
\begin{tabular}{|c|c|c|}\hline
Channel & Tree--Level & Penguin \\\hline
\hline
$B_c^\pm \to D_s^\pm a$		 &	$ (6-160)\times 10^{-11}$ 		& 	$ 2 \times10^{-6}$  			\\\hline
$B_c^\pm \to D^\pm a$	        	 &       $ (1-30)\times 10^{-11} $		 &	$ 3 \times10^{-7}$				\\\hline
$B_c^\pm \to K^{*\pm} a$         	 &        $ (2-70)\times 10^{-11} $		 &	n.a.							\\\hline
$B_c^\pm \to \rho^\pm a$           	 &         $ (4-100)\times 10^{-11} $ 		&	n.a.							\\\hline
$B_c^\pm \to K^\pm a$               	 &        $ (8-230)\times 10^{-12}$		&	n.a.							 \\\hline
$B_c^\pm \to \pi^\pm a$			 &	 $(3-85)\times10^{-11}$		&	n.a.							\\\hline\hline
$B^\pm \to D_s^\pm a$			 &	$ (5-30)\times 10^{-12}$ 		& 	n.a.							 	\\\hline
$B^\pm \to D^\pm a$	          	 &    $ (1-7)\times 10^{-12}$		 	&	n.a.					\\\hline
$B^\pm \to K^{*\pm} a$	          	 &        $ (1-7) \times 10^{-12}$		&	 $4\times 10^{-6}$		\\\hline
$B^\pm \to \rho^\pm a$	          	 &       $ (3-20)\times 10^{-12} $ 		&	  $4\times 10^{-7}$			\\\hline
$B^\pm \to K^\pm a$              	&       $(8-50)\times 10^{-13} $		&	 $2\times 10^{-6}$		 \\\hline
$B^\pm \to \pi^\pm a$			&	 $(3-20)\times10^{-12}$			&	 $3 \times 10^{-7}$			 	\\\hline\hline
$D_s^\pm \to K^{*\pm} a$		&        $ (1-60) \times 10^{-11}$ 		&	  $ 6 \times 10^{-12}$ 		\\\hline
$D_s^\pm \to \rho^\pm a$		&        $ (3-170) \times 10^{-11}$		&	n.a.							\\\hline
$D_s^\pm \to K^\pm a$	        	&        $ (6-300) \times 10^{-12} $ 		&	 $7\times 10^{-12}$ 			\\\hline
$D_s^\pm \to \pi^\pm a$            	&        $ (2-120) \times 10^{-11} $		&	n.a.						 \\\hline\hline
$D^\pm \to K^{*\pm} a$	 		&        $ (2-100) \times 10^{-12}$ 		&	n.a.							\\\hline
$D^\pm \to \rho^\pm a$	 		&        $ (7-290) \times 10^{-12}$ 		&	$3\times 10^{-12}$			\\\hline
$D^\pm \to K^\pm a$	    	        &        $ (1-50) \times 10^{-12} $ 		&	n.a.						\\\hline
$D^\pm \to \pi^\pm a$           		&        $ (5-200) \times 10^{-12}$  		&	$6\times 10^{-12}$		 \\\hline\hline
$K^{*\pm} \to K^\pm a$            	&        $ (5-25)\times 10^{-13} $	 	&	$4\times 10^{-8}$		 \\\hline
$K^{*\pm} \to \pi^\pm a$          	&        $ (3-20) \times 10^{-12}$ 		&	$3\times 10^{-9}$	 \\\hline\hline
$\rho^{\pm} \to K^\pm a$            &        $ (8-25)\times 10^{-12} $	 	&	$2\times 10^{-9}$	 \\\hline
$\rho^{\pm} \to \pi^\pm a$         	&        $ (3-9) \times 10^{-11}$ 		&	$4\times 10^{-10}$	 \\\hline\hline
$K^\pm \to \pi^\pm a$               	&        $ (2- 10) \times 10^{-12} $ 		&	$5\times 10^{-10}$	 \\\hline
\end{tabular}
\caption{Tree-level and penguin contribution to the hadronic charged meson decay rates, calculated for $m_a=0$, $f_a=1$ TeV and 
$c_f=\pm 1$, expressed in GeV${}^{-1}$ units. The interval in the tree--level column is obtained by setting $c_Q/c_q =(1,-1)$.}
\label{tab:constraints}
\end{table}

It is also useful to note from Eqs.~(\ref{eq:happrox_1}) and (\ref{eq:happrox_2}) that the ratio between the tree-level 
ISR and FSR amplitudes is always independent of the particular nature of the decay ($s$ or $t$) and, with the only exception 
of the vector to vector meson decay, is given by:
\beq
R^T_{I/F} = \left| \frac{\mathcal{M}^{(s,t)}_{I}}{\mathcal{M}^{(s,t)}_{F}} \right| \simeq \left(\frac{M_I}{M_F}\right)^2,
\eeq
with the identity strictly holding in the ALP massless limit and assuming a ``very light'' or a ``very heavy'' DA function. 
Therefore, for all the corresponding processes, the decay amplitude is always dominated by the ISR ALP emission. 
\begin{table}[t!]
\centering
\begin{tabular}{|c|c|c|}\hline
Channel & Tree--Level & Penguins \\\hline
\hline
$B_s^0 \to D_s^0 a$	     		& 	n.a.						&	$ 4 \times10^{-7}$					\\\hline
$B_s^0 \to D^0 a$	     			& $  (7-70)\times10^{-12}$		&	n.a.							\\\hline
$B_s^0 \to K^{*0} a$               	&         n.a.					&	 $4 \times10^{-6} $	 	\\\hline
$B_s^0 \to\rho^0 a$	   		&   $ (4-50)\times 10^{-13} $	& 	n.a.						\\\hline
$B_s^0 \to K_L^0 a$                	&   	n.a.						&	$ 3 \times 10^{-7}$  			 \\\hline\hline
$B^0 \to  K^{*0}   a$			&   	n.a.						&	 $ 4 \times10^{-6} $	 			\\\hline
$B^0\to D^0 a$				&	$ (3-30)\times 10^{-11}$ 		&   	n.a.								\\\hline
$B^0 \to  \rho^0 a$	           		&       $ (2-20)\times 10^{-12}$ 		& 	$6 \times 10^{-7} $		\\\hline
$B^0\to K_L^{0} a$	         	  	&      n.a.						&	 $4\times 10^{-6}$			\\\hline
$B^0\to \pi^0 a$		           	&       $ (1-10)\times 10^{-12} $ 	&	  $ 5 \times 10^{-7}$		\\\hline\hline
$D^0\to K^{*0} a$		           	&        $ (7-300) \times 10^{-12}$	& 	n.a.							\\\hline
$D^0\to \rho^0 a$		           	&        $ (5-200) \times 10^{-12}$	&	 $4\times 10^{-12}$			 \\\hline
$D^0\to K_L^0 a$		           	&        $ (7-270) \times 10^{-13}$	&	 n.a.					 	\\\hline
$D^0\to \pi^0 a$		           	&        $ (2-100) \times 10^{-12}$	&	 $3\times 10^{-12}$			\\\hline\hline
$K^{*0}\to K^0 a$		           	&        $ (2-6) \times 10^{-12}$		&	$3\times 10^{-9}$			\\\hline
$K^{*0}\to \pi^0 a$				&        $  (1-2) \times 10^{-11}$		&	 $ 3\times 10^{-9}$				\\\hline\hline
$\rho^{0}\to K^0 a$				&        $ (1-3) \times 10^{-11}$ 		&	 $ 2\times 10^{-9}$				\\\hline
$\rho^{0}\to \pi^0 a$				&        $ (2-7) \times 10^{-11}$		&	 $ 3\times 10^{-9}$			\\\hline\hline
$K_L^0\to \pi^0 a$			        &        $ (4-20) \times 10^{-15}$		&	 $1\times 10^{-10}$		\\\hline
\end{tabular}
\caption{Tree-level and penguin contribution to the hadronic neutral meson decay rates, calculated for $m_a=0$, $f_a=1$ TeV and 
$c_f=\pm 1$, expressed in GeV${}^{-1}$ units. The interval in the tree--level column is obtained by setting $c_Q/c_q =(1,-1)$.}
\label{tab:constraints2}
\end{table}
Finally, in Tab.~\ref{tab:constraints}, both the tree-level and penguin contributions, when available, are presented. As a rule 
of thumb the tree-level vs one-loop amplitudes ratio can be estimated by: 
\bea
R_{T/L} = \left| \frac{\mathcal{M}_\mathrm{T}^{(s,t)}}{\mathcal{\mathcal{M}_\mathrm{L}}} 
\right| \approx 2\,\pi^2 \frac{f_I \, f_F}{m_f^2} 
\left|\frac{V^\mathrm{CKM}_\mathrm{T}}{V^\mathrm{CKM}_\mathrm{L}} \right|,
\label{eq:treeloopratio}
\eea
where $m_f$ is the mass of the heaviest quark running in the penguin and $f_{I,F}$ the initial and final meson decay constants. Notice that 
for most of the $D$ decays the tree-level contribution is comparable if not larger than the loop one, as clearly the $m^2_b$ 
penguin loop enhancement is not sufficient to compensate for the typical loop suppression factor. Conversely, for the $K$ and 
$B$ meson sector the tree/loop ratio looks really tiny thanks to the large $m_t^2$ penguin enhancement. Nevertheless, for the $K$ 
sector the tree-level diagrams may have a non negligible impact, as they depend on different, and often less constrained, 
ALP-fermion couplings, as discussed in \cite{Guerrera:2021yss}.

In Tab.~\ref{tab:constraints2} the decay rates of several neutral meson decays are collected. For definiteness, again $m_a=0$, 
$f_a=1$ TeV and $c_f=\pm 1$ have been used. The same formula discussed for the charged meson decays can be straightforwardly 
obtained also for the neutral meson case. In particular, penguin amplitudes typically dominate over tree-level ones, when 
available. The only exception being again represented by the $D$ meson sector, where tree-level amplitudes are at least one 
order of magnitude larger than penguin ones. 

The Kaon sector is the sector from which the most precise bounds on ALP-fermion couplings are obtained \cite{Izaguirre:2016dfi,
Gavela:2019wzg,Guerrera:2021yss}, thanks to very precise decay rate measurements. In particular NA62 at CERN, looking at 
$K^+\to \pi^+ \,a$, has collected $3\times 10^{16}$ p.o.t. in Run 1 and is aiming for $10^{18}$ p.o.t by the end of Run 2. 
Using the complete Run 1 dataset, the NA62 experiment established an upper limit on the $\mathcal{B}(K^+\to \pi^+\,a)$ for an 
invisible ALP at the level of $10^{-11}$ in the mass ranges of 0--110 MeV and 155--260 MeV \cite{CortinaGil:2021nts,
CortinaGil:2020fcx}. NA62 experiment has also established upper limits on $\mathcal{B}(K^+ \to\pi^+\,a) \lesssim 10^{-9}$ in 
the 110--115 MeV mass range, i.e. around of the $\pi_0$ mass, from a dedicated analysis based on the 10\% of the Run 1 minimum 
bias dataset \cite{NA62:2020pwi}. Measurements of the $K^0_L\to \pi^0 \bar \nu \nu$ decay naturally provide limits on the 
$\mathcal{B}\left( K^0_L\to \pi^0 a\right)$ branching ratio. KOTO experiment \cite{Ahn:2018mvc} has reported a limit on 
$\mathcal{B}(K^0_L \to \pi^0 \,a) \lesssim 2.4\times 10^{-9}$ at 90\% CL with the 2015 dataset, practically independent on the ALP 
mass up to the kinematical limit. 

\begin{figure}[ht!]
\centering \hspace{-1.2cm}
\includegraphics[scale=0.8]{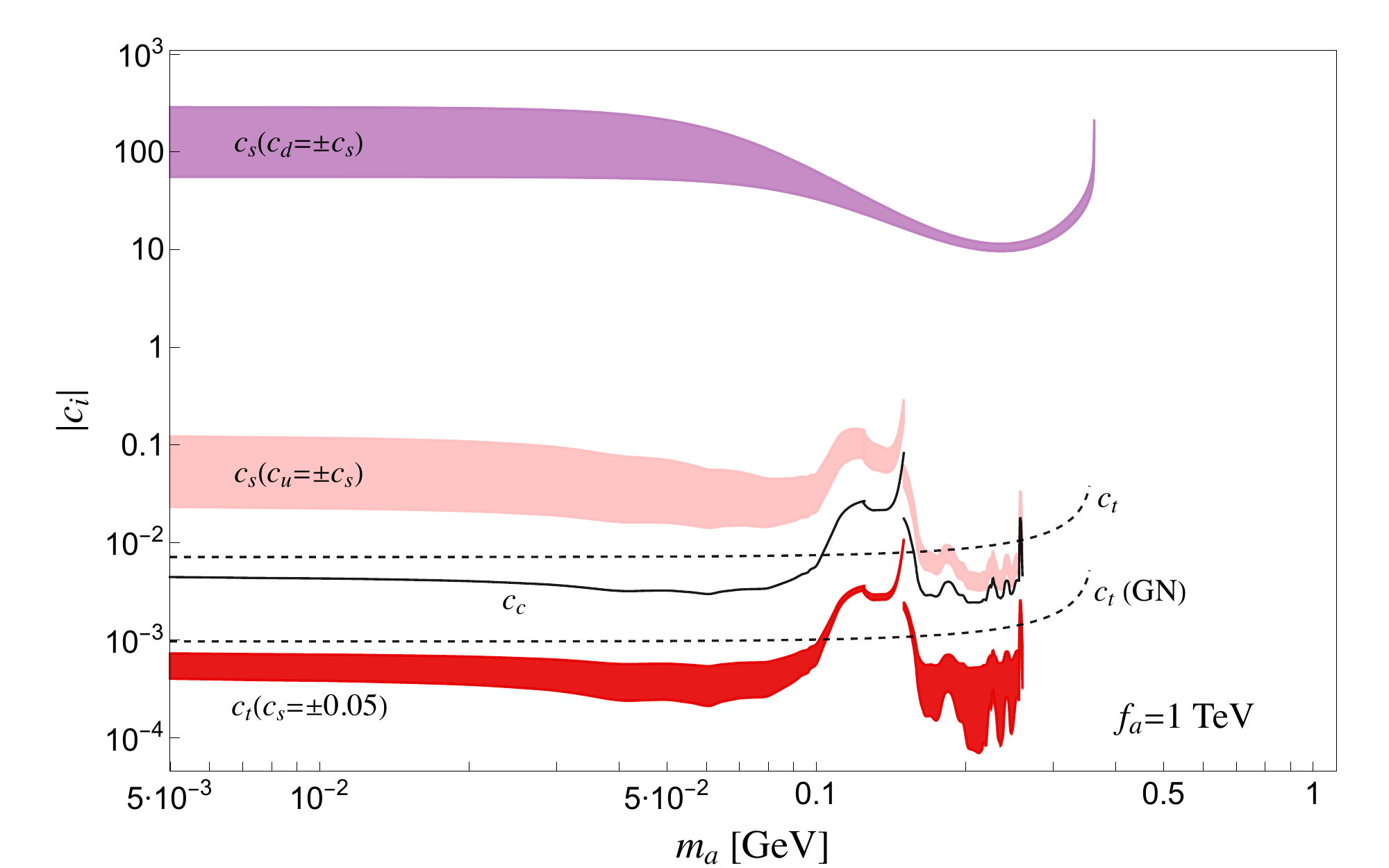}
\caption{Limits on ALP-quark couplings from hadronic $K$ meson decays, as function of the ALP mass $m_a$ for $f_a=1$ TeV. 
The meaning of the different lines/regions is explained in the text.}
\label{fig:sum_plota}
\end{figure}

\begin{figure}[h!]
\centering \hspace{-1.2cm}
\includegraphics[scale=0.8]{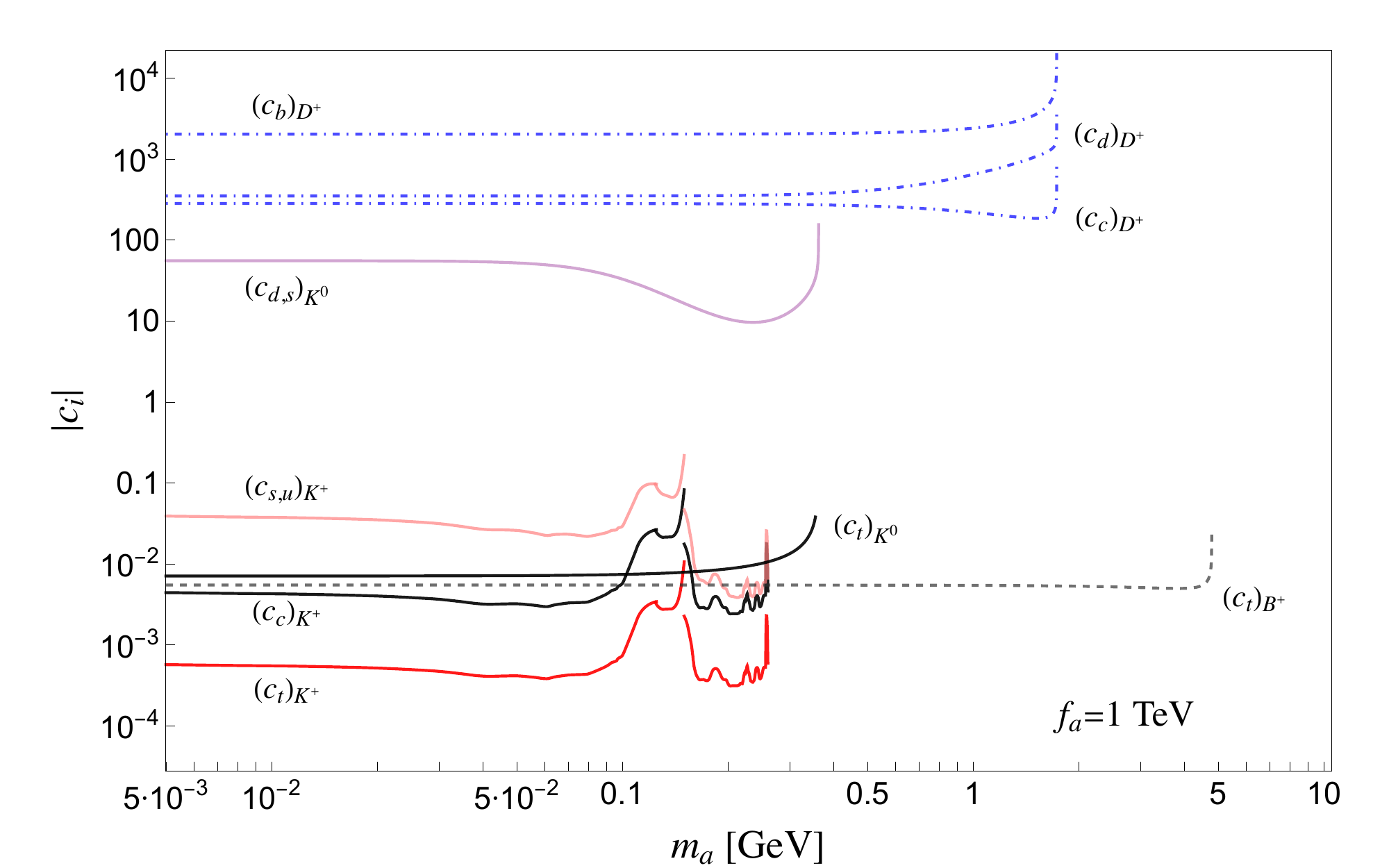}
\caption{Limits on ALP-quark couplings from hadronic meson decays, as function of the ALP mass $m_a$ for $f_a=1$ TeV. The meaning 
of the different lines is explained in the text.}
\label{fig:sum_plotb}
\end{figure}

In Fig.~\ref{fig:sum_plota} a summary of the constraints on the quark couplings from $K$ hadronic decays into an invisible ALP 
are collected as function of the mass $m_a$ and for the chosen reference value $f_a=1$ TeV. The pink (middle) and red (lower) 
shaded areas show the bounds directly derived in Ref.~\cite{Guerrera:2021yss} from the $K^+\to\pi^+\,a$ decay rate. The pink area 
represents the limits projected onto the ALP-valence s- and u-quarks couplings, assuming all the other ALP-fermion coupling 
vanishing. Given that the tree-level amplitude is largely dominated by the ISR amplitude, it depends mainly on $c_s$ and $c_u$. 
The highlighted region has a meaning similar to the range displayed in Tab.~\ref{tab:constraints} and~\ref{tab:constraints2}, 
i.e. the boundaries of the region correspond to two the limiting cases $c_u=\pm c_s$. The single parameter limit on $c_{s(u)}$, 
obtained by setting $c_{u(s)}=0$, lies approximately in the middle of the allowed range. The red (lower) shaded area, instead, 
represents the bound on $c_t$ obtained by NA62 data assuming a non vanishing $c_t$ coupling and letting $c_s$ vary in the range 
$[-0.05,0.05]$. Therefore, despite the fact that the penguin contribution to the $K^+\to\pi^+\, a$ decay is two order of magnitude 
larger than the tree-level one, a contamination of the $c_t$ coupling of roughly one order of magnitude is still possible\footnote{See 
Ref.~\cite{Guerrera:2021yss} for a detailed discussion on the ALP-fermion couplings bounds from $K\to\pi\, a$ decays.}. The single 
parameter limit on $c_{t}$, obtained by setting all other $c_i=0$, lies approximately in the middle of the allowed range. The 
continuous middle black line shows the single parameter limit on $c_c$, coupling associated to the sub-dominant term weighting 
roughly 10\% of the total penguin contribution. The upper black dashed line represents the exclusion limit on the $c_t$ parameter, 
obtained from the KOTO $K^0_L\to\pi^0\nu\bar{\nu}$ measurements, while the lower black dashed one represents the exclusion limit on 
$c_t$ obtained using NA62 $K^+\to\pi^+\nu\bar{\nu}$ branching ratio for inferring a limit on the $K^0_L$ branching ratio through the 
Grossman-Nir bound. As the top penguin loop largely dominate the $K^0_L$ decay rate, see Tab.~\ref{tab:constraints2}, no appreciable 
contamination from the tree-level diagrams is expected. Finally the (upper) violet shaded area represents the limits projected 
onto the ALP-valence s- and d-quarks couplings from the KOTO $K^0_L$ measurement, assuming all the other ALP-fermion coupling 
vanishing and corresponding to the two limiting cases $c_d=\pm c_s$. Therefore, in Fig.~\ref{fig:sum_plota} an exhaustive set 
of bounds on ALP-quark couplings obtained from hadronic $K$ decays is collected. The single parameter limits obtained from 
hadronic $K$ decays are also reported in Fig.~\ref{fig:sum_plotb} as continuous colored lines, for comparison.

Hadronic $B$ meson decays in ALP are expected to provide additional interesting bounds on ALP-fermion couplings. Still not 
competing with the ones extracted from $K$ decays, these bounds nevertheless suffer from a smaller theoretical uncertainty 
and extend the $m_a$ range of constraints up to a few GeV. The first thing one can notice from Tab.~\ref{tab:constraints} and 
Tab.~\ref{tab:constraints2} is that, for the $B$ meson sector, the tree-level vs loop ratio is two/three order smaller compared 
to the $K$ meson ratio, mainly due to the larger CKM suppression factor. Therefore, one expects practically no contamination on the 
$c_t$ bounds from ALP-fermion couplings entering in the tree-level amplitudes, assuming perturbativity. The lower dashed black line 
in Fig.~\ref{fig:sum_plotb} is the $c_t$ exclusion bounds obtained from the recent $B\to K a$ Belle II data \cite{Dattola:2021cmw}. 
Limits on $B$ valence quark couplings are extremely weak due to the smallness of the tree-level amplitude and lie in the 
$(10^4-10^5)$ region and are not shown in figure\footnote{Tree level contributions of $B$ meson valence quark 
 may have possible interferences with initial/final state emission, as commented in Sec.~\ref{subs:penguin}.}.
 From Tab.~\ref{tab:constraints} one can identify another very promising channel, namely $B_c^+\to D_s^+\,a$. 
A future experimental limit on $\mathrm{Br}(B_c^+\to D_s^+ a)<10^{-5}$ would provide competitive bounds on $c_t$. Other 
theoretically interesting channels are hadronic $B_c$ decays into vector meson, like $B_c^\pm\to \rho (K^*)\, a$. These decays 
have no penguin contribution and a small $M_{\rho,K^*}/M_{B_c}$ mass ratio. Therefore a very clean extraction on ALP-$B_c$ 
valence quark coupling could be addressed. 

A complementary set of information can be potentially extracted from hadronic $D$ decays in ALP. Indeed, $D$--mesons penguins do 
not dominate anymore over tree-level amplitudes being $m_b/m_t$ suppressed compared to similar $B$ and $K$ meson decays, as shown 
numerically by Tab.~\ref{tab:constraints} and Tab.~\ref{tab:constraints2}. At present there are no experimental results measuring 
hadronic $D$ decays in ALPs, the only information originate from a recast of the charged $D\to \tau(\to \pi \nu)\bar{\nu}$ decay 
onto the $D^\pm\to \pi^\pm \, a$ branching ratio, as proposed by Ref.~\cite{MartinCamalich:2020dfe}. Even though the predicted 
signals are quite weak these channels can provide sensitivity on the $D^\pm$ valence quark couplings, $c_d$ and $c_c$ and 
eventually on $c_b$ through the dominant down-type penguin loop. The individual limits on these ALP-quark couplings, shown 
in Fig.~\ref{fig:sum_plotb} as dot-dashed lines, appear evidently above the perturbativity region once $f_a=1$ TeV is chosen. 

The single parameter limits obtained from hadronic $K$ meson decays are represented as continuous colored lines. Finally, no 
experimental data are at present available for $K^*$ and $\rho$ hadronic decays in ALP. The expected pattern is assumed to be, 
however, very similar to the $K$s one.


\subsection{Leptonic Final States}
\label{sec:Leptonic_Final_States}

Pseudoscalar leptonic decay can be used to constraint flavor–diagonal ALP-fermion couplings through the tree-level 
amplitudes of Eqs.~(\ref{eq:MMALP}) and (\ref{eq:MLALP}). In the case of invisible ALP, considered throughout this paper, the 
simplest approach is to saturate the 1-$\sigma$ experimental limits on the corresponding SM leptonic branching ratio adding the 
pseudoscalar meson three-body leptonic ALP decay to the two-body leptonic SM one, having the same missing energy signature. At 
present, in fact, there is not enough available experimental information on the charged lepton energy distribution such that 
one can obtain stricter bounds by characterizing two-body vs three-body decay spectrum\footnote{See \cite{Gallo:2021ame} 
for more details on three-body spectral analysis of meson leptonic decays in ALP.}.

Leptonic $B$ decays have been measured at Babar and Belle. The latest Belle data for electron, muon and tau channel can be found in 
\cite{Belle:2006tbq,Belle:2019iji,Belle:2012egh}, respectively. Charmed meson decays have been measured at BESS (see 
\cite{Eisenstein:2008aa,BESIII:2013iro,BESIII:2018hhz} for $D$ and~\cite{BESIII:2016cws,BESIII:2019vhn} for $D_s$ decays respectively) 
and at Belle \cite{Belle:2013isi}. Leptonic Kaon decays have been measured by KLOE and NA62~\cite{NA62:2012lny,KLOE:2007wlh,
ParticleDataGroup:2020ssz}. 

\begin{table}[t]
\centering
\begin{tabular}{|c||c||c||c|}\hline
Channel							& $c_i$ u-type  			& 		$c_i$ d-type 	&	$c_i$ leptons		\\ \hline\hline
$B^\pm\!\! \to\! e^\pm \bar{\nu}_e$		&	360				&	250				&	2$\cdot 10^{6}$		\\\hline
$B^\pm\!\! \to\! \mu^\pm \bar{\nu}_\mu$	&	170				&	120				&	5$\cdot 10^{3}$			\\\hline
$B^\pm\! \!\to \!\tau^\pm \bar{\nu}_\tau$	&	2.6$\cdot 10^{3}$	&	2$\cdot 10^{3}$		&	5$\cdot 10^{3}$		\\\hline\hline
$D^\pm\!\! \to \!e^\pm \bar{\nu}_e$		&	170				&	175				&	5$\cdot 10^{4}$	 	 \\\hline
$D^\pm\! \!\to\! \mu^\pm \bar{\nu}_\mu$	&	250				&	270				&	3.5$\cdot 10^{3}$	\\\hline
$D^\pm\! \!\to \!\tau^\pm \bar{\nu}_\tau$	&	1.5$\cdot 10^{5}$	&	1.5$\cdot 10^{5}$	&	1.7$\cdot 10^{5}$	\\\hline\hline
$D_s^\pm\!\! \to\! e^\pm \bar{\nu}_e$		&	125				&	120				&	5$\cdot 10^{5}$		 \\\hline
$D_s^\pm\!\! \to\! \mu^\pm \bar{\nu}_\mu$	&	180				&	175				&	3.5$\cdot 10^{3}$	\\\hline
$D_s^\pm \!\! \to \!\tau^\pm \bar{\nu}_\tau$&	5$\cdot 10^{4}$		&	1$\cdot 10^{5}$		&	5$\cdot 10^{4}$	 	 \\\hline\hline
$K^\pm \!\! \to\! e^\pm \bar{\nu}_e$		&	4				&	6				&	4$\cdot 10^{3}$		\\\hline
$K^\pm \!\! \to \!\mu^\pm \bar{\nu}_\mu$	&  	600	 		 	&	800				&	3$\cdot 10^{3}$	\\\hline
\end{tabular}
\caption{Limits on single ALP-fermion coupling $c_i$ derived from pseudoscalar meson leptonic decays for $m_a=0$ and $f_a=1$ TeV.}
\label{tabfa}
\end{table}
The derived bounds on single ALP-fermion couplings, $c_i$, for $m_a=0$ and $f_a=1$ TeV are shown in Tab.~\ref{tabfa}. As an example, 
the first row in Tab.~\ref{tabfa} should be read as follows: the ``up–quark'' column represents the limit on $c_u$ by setting $c_b = 
c_e = 0$, the ``down–quark'' column represents the limit on $c_b$ by setting $c_u = c_e = 0$, and finally the value in the ``lepton'' 
column is the limit on $c_e$ for $c_u = c_b = 0$. It is true in general that hadronic meson decays in ALP can provide by far the most 
stringent limits on a universal ALP-fermion coupling, bounding $c_{a\Phi} \lesssim 5 \times 10^{-4}$ from $K$ decays or $c_{a\Phi} 
\lesssim 8 \times 10^{-3}$ from $B$ decays, for $f_a=1$ TeV. However, both these limits come associated to the top enhanced penguin, 
and therefore in a non-universal ALP-fermion coupling scenario can be applied to bound solely $c_t$. On the contrary, leptonic and 
hadronic decays in ALP often provide similar constrains on single ALP-light-quark couplings, as can be seen comparing the results 
of Tab.~\ref{tabfa} with the ones summarized in Fig.~\ref{fig:sum_plotb}. 
While, for example, single parameter bounds on $c_{u,s}$ derived from charged $K$ hadronic decay are still one order of magnitude 
better than the ones obtained from charged $K$ leptonic channel, conversely $c_c$ bounds obtained from charged $D$ meson leptonic 
decay are of the same order, or even slightly better than the corresponding charged $D$ meson hadronic ones. Finally, $c_b$ bounds 
from charged $B$ meson leptonic decays are at least two order of magnitude better than the hadronic limits. Therefore, meson 
leptonic decays provide useful complementary information once independent bounds on all the ALP-quark couplings are needed. 

\begin{figure}[t!]
\center
\includegraphics[scale=0.4]{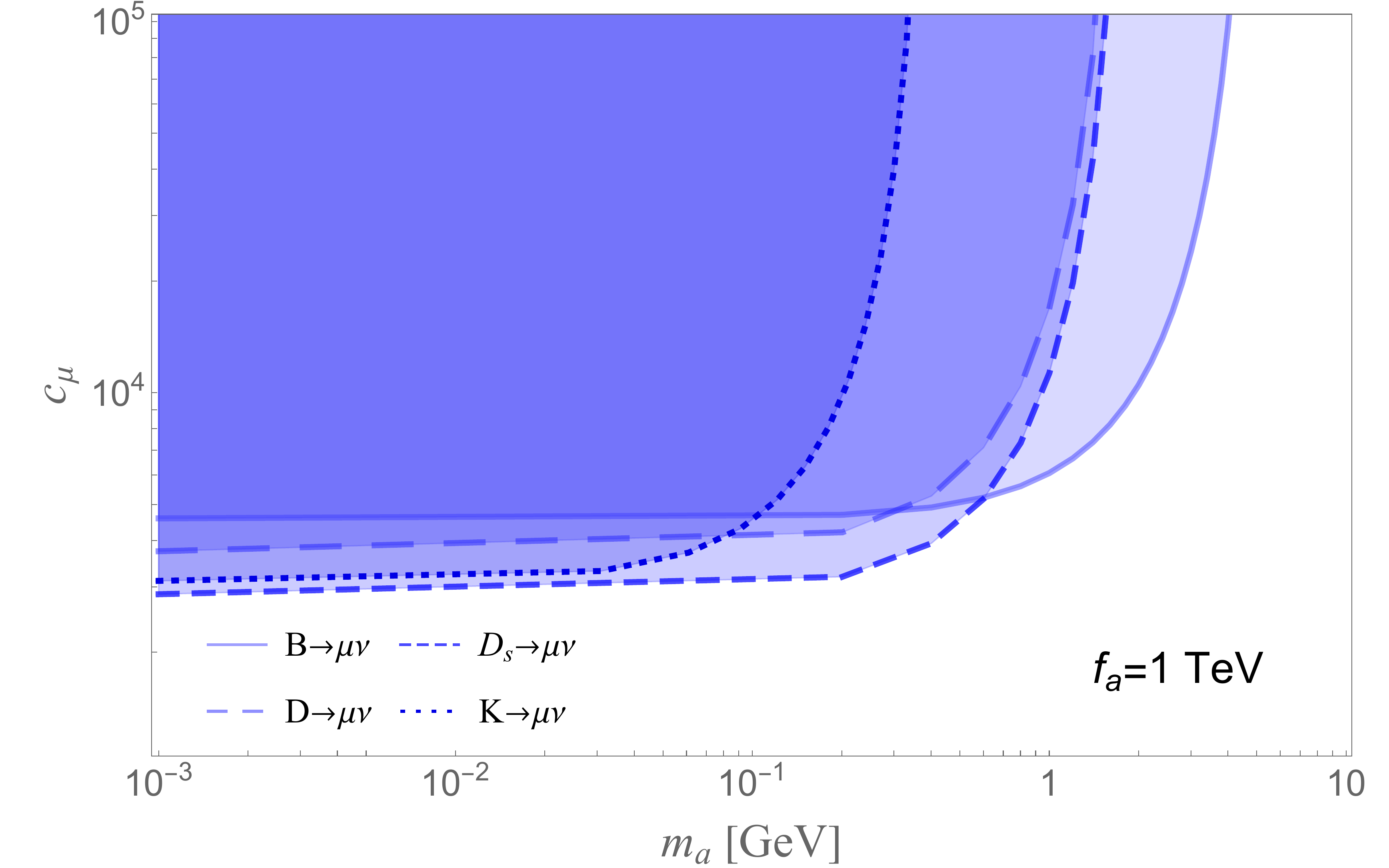}
\caption{Summary of the limits extracted from the semileptonic class, $M\to\ell \nu a$, on the muonic coupling $c_\mu$ with $f_a$ fixed at 1 TeV.\label{fig:semileptonic_summ}}
\end{figure}

From Tab.~\ref{tabfa}, evidently emerges that limits on ALP-lepton coupling are very weak, due to the combination of low masses 
(i.e. the electron case) and/or not very precise experimental results (i.e. the $\tau$ channels). However, despite these 
feeble bounds, pseudoscalar meson leptonic decays in ALP provide undoubtedly the best available limits on the ALP–lepton sector 
for an $m_a$ in the KeV-GeV range. In Fig.~\ref{fig:semileptonic_summ}, for exemplification, all the limits on the muon 
coupling, $c_\mu$, are collected as function of the ALP mass $m_a$, for the chosen values $f_a=1$ TeV. From the plot one can 
easily discern a slower saturation of the kinematical limit compared to the corresponding hadronic decays. Already at values 
$m_a \approx M_M/2$ a strong reduction of the decay rate appears. One can easily understand from the corresponding Dalitz plot 
that this effect is associated to the tree-body nature of the leptonic meson decay in ALP. Therefore some additional caution 
should be used in this case to generalize the validity of $m_a=0$ results to higher mass values.

\subsection{Phenomenological Summary}
\label{sec:phenomenology}

\begin{figure}[t!]
\center
\includegraphics[scale=0.43]{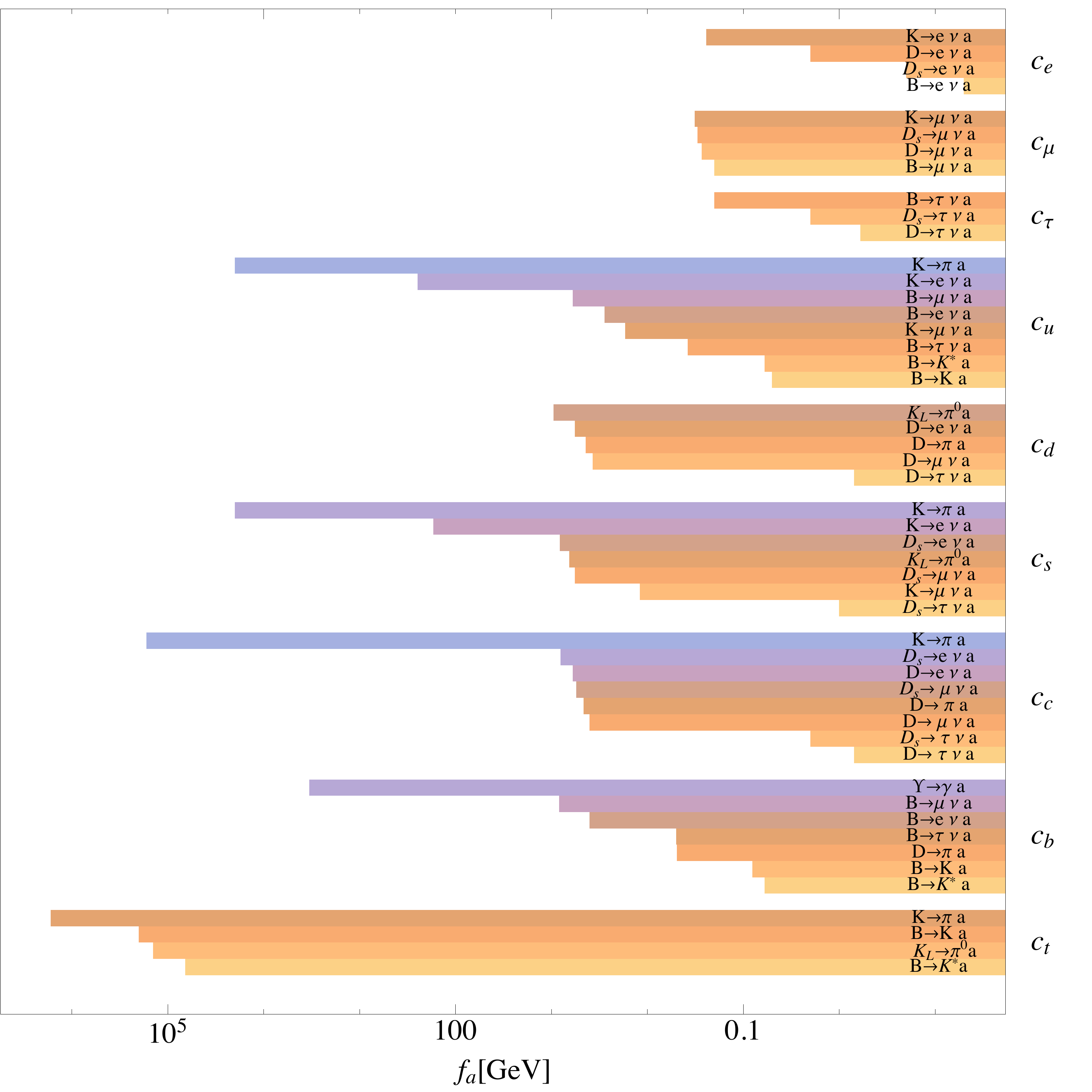}
\caption{Bounds on $f_a$ (expressed in GeV) obtained from meson decays into an invisible ALP, for $m_a=0$ and assuming the 
corresponding $c_i=1$, while setting all the other couplings to $0$.} 
\label{final_summary}
\end{figure}
Finally, a comprehensive summary of all the bounds on flavor conserving ALP--fermion couplings derived in the previous section is 
presented in Fig.~\ref{final_summary}. To be able to fairly compare all the different analysis, the limits on the $U(1)_{PQ}$ breaking
scale $f_a$ (expressed in GeV) are shown, for $m_a=0$ and by assuming the corresponding $c_i=1$, with all the other couplings 
set to 0. Therefore, the $f_a$ value plotted represents the highest energy scale tested, at present, in each decay channel. 
One can notice that the most stringent bounds on $f_a$ come from the top sector, trough the $m_t$ enhanced penguin contributions 
\cite{Gavela:2019wzg,Guerrera:2021yss}. From $K$ and $B$ hadronic pseudoscalar meson decays one tests $f_a \simeq (10^5-10^6)$ 
GeV. This is the lowest energy scale at which new physics in the ALP sector may appear, when a universal ALP-fermion coupling 
is assumed. $K\to \pi\,a$ decay provides the strongest bounds for all the ALP-quark couplings in the non-universal, but flavor 
conserving ALP-fermion scenario, with the only exception of the ALP-bottom coupling where the strongest bound comes from 
the $\Upsilon(ns)\to \gamma \,a$ decay \cite{Merlo:2019anv}. Bounds on $c_{t,c}$ enter from the penguin loop diagram, while 
bounds on $c_{u,d,s}$ are due to the tree-level amplitude contribution. Meson leptonic decays produces an upper bound on 
$f_a \simeq (10^{-1}-10)$ GeV in the ALP-lepton sector.


\section{Conclusions}

In this paper, bounds on non-universal flavor--diagonal ALP--fermion couplings are presented. These limits have been extracted 
from mesonic decays assuming an invisible ALP signature. Two large classes of processes are studied: i) hadronic meson decays 
$M_I\to M_F \,a$, with $M_I$ and $M_F$ pseudoscalar and/or vector mesons and ii) leptonic meson decays $M\to \ell\nu a$. 
Lattice QCD and Brodsky--Lepage method are used for calculating the hadronic matrix elements associated to local and bi--local 
operators respectively. In particular, a complete set of tree-level amplitudes intervening in mesonic ALP decays have been derived 
for the first time, allowing a general comparison between tree-level and penguin mediated processes. For example, hadronic $K$ 
and $B$ decays in ALP are both clearly top-penguin dominated. However, already from a quick analysis of Eq.~(\ref{eq:treeloopratio}) 
one can estimate the different level of parameter cross-contamination that one could expect in a general non universal flavor 
conserving framework: while can be sizable for the case of $K$ meson hadronic decays, as exemplified in Fig.~\ref{fig:sum_plota}, 
it is instead completely negligible for the $B$ sector as explained in the text and shown in Fig.~\ref{final_summary}. Conversely, 
hadronic $D$ meson decays are typically tree-level dominated but a large penguin contamination is however expected. 
Moreover, a complete analysis of the independent limits on the different ALP-charged lepton diagonal couplings 
has been presented. Despite being several order of magnitude less constrained that the quark counterpart, these bounds extend 
previous limits, obtained mainly from astrophysical data, to the KeV-GeV ALP mass range. Finally, for completeness in 
App.~\ref{sec:FlavourViolating} a recast of the bounds in term of general, non-diagonal, ALP-fermion couplings is presented.  

The work presented here can be easily extended to visible ALP decays. It is however phenomenologically much more complicated 
to use those data for performing model independent analysis, as each decay amplitude will depend on the combinations of 
products of two different ALP-fermion couplings.


\section{Acknowledgements}
We thank Xavier Ponce D\'iaz for useful discussion. A.G. and S.R. acknowledge support from the European Union’s Horizon 2020 
research and innovation programme under the Marie Sklodowska-Curiegrant agreements 690575 (RISE InvisiblesPlus), 
674896 (ITN ELUSIVES) and 860881 (HIDDEN).

\appendix
\section{Limits on Flavor Violating Couplings}
\label{sec:FlavourViolating}

For the sake of completeness a projection of the limits induced onto the flavor changing parameters of the Lagrangian in 
Eq.~(\ref{eq:General_L}), is presented here. The dimension five 
ALP-quark Lagrangian reads:
\beq
\mathcal{L}^a_\mathrm{FC}=\frac{\partial_\mu a}{2f_a} \bar{d}\gamma^\mu(C^{(d)}_V+C^{(d)}_A\gamma^5)d+ 
\frac{\partial_\mu a}{2f_a} \bar{u}\gamma^\mu(C^{(u)}_V+C^{(u)}_A\gamma^5)u.
\label{eq:General_flavor_violating}
\eeq
$V$-type couplings will induce parity conserving mesonic decays such as $\PS_I\to \PS_F\, a$ and $\VE_I\to \VE_F\, a$, while  
$A$-type one will enter in parity violating processes such as $\PS\to \VE\, a$. 

Two different scenarios may be considered: 1) flavor violation is induced by tree level parameters or 2) flavor violation is 
induced by effective couplings that emerge due to RG effects.  In the first scenario, the $M_I\to M_F \,a$ branching ratio, 
in the ALP massless limit, reads:
\beq
\mathrm{Br}(M_I\to M_F \,a) = \kappa\frac{M_I^3}{16\pi\Gamma_I}\frac{\left|\left( C^{(u,d)}_{A,V}\right)_{ij}\right|^2}
{4 f^2_a}F(m_a^2)^2\left(1-\frac{M^2_F}{M^2_I}\right)^3,
\eeq
where $(C^{(q)}_{A,V})_{ij}$ is the relevant coupling mediating the FC transition and $F(m_a^2)$ is the associated hadronic form factor. 
The parameter $\kappa$ is 1 (1/3) if $M_I$ is a pseudoscalar (vector) meson respectively. The bounds extracted from hadronic meson 
decays into an invisible ALP are collected in Tab.~\ref{tab:fv_constraints}, for the chosen value $f_a=1$ TeV. From $K^+ \to \pi^+ \,a$ 
decay one can test the $sd$-vector sector, $(C^{(d)}_V)_{sd}$. To bound the $sd$-axial sector, $(C^{(d)}_A)_{sd}$, one could use 
the $K^* \to \pi\,a$ decays. These decays, however, are not measured yet, and therefore such limits have to 
be expressed as function of a still unknown branching ratio (Br):
\beq
\Big|(C^{(d)}_A)_{sd}\Big|=6 \cdot 10^3\cdot\left[\mathrm{Br}(K^* \to \pi\,a)\right]^{1/2}\, \mathrm{GeV}^{-1}.\label{eq:a1}
\eeq 
The $B^+\to K^+ a$ and $B^+\to K^{+*} a$ channels provide exclusion limits on $(C^{(d)}_V)_{bs}$ and $(C^{(d)}_A)_{bs}$ respectively.
Finally $(C^{(d)}_V)_{bd}$ and $(C^{(d)}_A)_{bd}$ can be tested via $B\to \pi(\rho) a$ decays. Again, as the  $B\to \rho \cancel{E}$ 
branching ratio is not measured one can express the bound as 
\beq
\Big|(C^{(d)}_A)_{bd}\Big|=7.4 \cdot 10 ^{-4}\cdot \left[\mathrm{Br}(B\to \rho \,a)\right]^{1/2}\, \mathrm{GeV}^{-1}\,.\label{eq:a2}
\eeq 
The numerical difference between Eq.~(\ref{eq:a1}) and (\ref{eq:a2}) is due to the huge difference in 
 the mean life of the resonances.
In the up-quark sector only the $cu$ sector can be tested via 
$D^+\to \pi^+(\rho^+) a$, yet not measured. Therefore, the limit on $(C^{(u)}_A)_{cu}$ is once again expressed as: 
\beq
\Big|(C^{(u)}_A)_{cu}\Big|=4.5 \cdot 10 ^{-3}\cdot \left[\mathrm{Br}(D\to \rho \,a)\right]^{1/2}\, \mathrm{GeV}^{-1}.
\eeq 
However, a limit on $(C^{(u)}_V)_{cu}$ can be extracted following Ref.~\cite{MartinCamalich:2020dfe} and using a recast of 
$D^+\to \tau^+ (\to \pi^+ \nu) \bar{\nu}$ giving $\mathrm{Br}(D^+\to\pi^+ a)<8\cdot 10^{-6}$. Processes involving top quark 
transitions are clearly not yet accessible.
\begin{table}[h!]
\centering
\begin{tabular}{| c | c | c | c |}\hline
	Vector 				&				 Limit 			& 		Axial				 & Limit  \\\hline
\hline
	$|(C^{(d)}_V)_{sd}|/f_a$		&	$2.5\cdot 10^{-12}$ GeV$^{-1}$ 	&	$|(C^{(d)}_A)_{sd}|/f_a$		&	n.a.						\\\hline
	$|(C^{(d)}_V)_{bs}|/f_a$		&	$9\cdot 10^{-9}$ GeV$^{-1}$ 		&	$|(C^{(d)}_A)_{bs}|/f_a$		&	$1.3\cdot 10^{-8}$ GeV$^{-1}$ \\\hline
	$|(C^{(d)}_V)_{bd}|/f_a$		&	$1\cdot 10^{-8}$ GeV$^{-1}$ 		&	$|(C^{(d)}_A)_{bd}|/f_a$		&	n.a. 						\\\hline
	$|(C^{(u)}_V)_{cu}|/f_a$		&	$2\cdot 10^{-8}$ GeV$^{-1}$ 		&	$|(C^{(u)}_A)_{cu}|/f_a$		&	n.a. 						\\\hline

\end{tabular}
\caption{Limits on flavor violating couplings in the scenario 1) for $m_a=0$ and $f_a=1$ TeV.}
\label{tab:fv_constraints}
\end{table}

In the second considered scenario one can set to 0 the off--diagonal parameters of Eq.~(\ref{eq:General_flavor_violating}) at the high 
scale $f_a$. This procedure does not get rid completely of flavor--violation in the ALP--sector, as it can be generated from the flavor 
conserving parameters via RG equations proportionally the SM flavor violation induced by the CMK matrix~\cite{Bauer:2020jbp}.
One can project limits on these effective couplings by equating the amplitude of a given process obtained from 
Eq.~(\ref{eq:General_flavor_violating}) to the amplitudes in Eq.~(\ref{loopAmplitudeK+}) and (\ref{loopAmplitudeVector}) for $V/A$ type 
couplings respectively. The off--diagonal entries of the matrices are to be considered not as tree level parameters, but as effective 
ones, induced by diagonal couplings. In principle there are four different matrices in the Lagrangian in 
Eq.~(\ref{eq:General_flavor_violating}), but this scenario does not distinguishes axial or vector off--diagonal elements. 
The effective flavor violating couplings read:
\beq
C^{(d/u)}_{ij}=\frac{G_F m_q^2}{2\sqrt{2}\pi^2}\sum_f c^{(f)}_{ij}=\frac{G_F m_q^2}{2\sqrt{2}\pi^2}\sum_f V_{fi}V_{fj}^*c_f \frac{x_f}{x_q}\ln\left(\frac{f_a^2}{m_f^2} \right)
\eeq 
where the sum over the $f$ runs on up-type quarks for $C^{(d)}$ and on down-type quarks for $C^{(u)},$ and $m_q$ is the mass of the heaviest 
quark running in the loop.  The limits shown in 
Fig.~\ref{final_summary}, recasted onto bounds on $C^{(d/u)}$, are reported in Tab.~\ref{tab:fv_constraints_2}, for $f_a=1$ TeV.
\begin{table}[h!]
\centering
\begin{tabular}{| c | c | c | c |}\hline
	$d$-type 				&				 Limit 			& 		$u$-type			 & Limit 						 \\\hline
\hline
	$|(C^{(d)})_{sd}|/f_a$		&	$3.8\cdot 10^{-12}$ GeV$^{-1}$ 	&	$|(C^{(u)})_{cu}|/f_a$		&	$3.2\cdot 10^{-12}$ GeV$^{-1}$  \\\hline
	$|(C^{(d)})_{bs}|/f_a$		&	$4.3\cdot 10^{-10}$ GeV$^{-1}$ 	&	$|(C^{(u)})_{tc}|/f_a$		&	$9.1\cdot 10^{-10}$ GeV$^{-1}$	  \\\hline
	$|(C^{(d)})_{bd}|/f_a$	&	$9.3\cdot 10^{-11}$ GeV$^{-1}$ 	&	$|(C^{(u)})_{tu}|/f_a$		&	$7.7\cdot 10^{-11}$ GeV$^{-1}$	   \\\hline
\end{tabular}
\caption{Limits on flavor violating couplings in the scenario 2) for $m_a=0$ and $f_a=1$ TeV.}
\label{tab:fv_constraints_2}
\end{table}
%

%


\bibliographystyle{hunsrt}
\bibliography{Tesi_Biblio}


\end{document}